# The Origination and Distribution of Money Market Instruments: Sterling Bills of Exchange during the First Globalisation

Olivier Accominotti
*London School of Economics and CEPR*

Delio Lucena-Piquero
*University of Toulouse, LEREPS*

Stefano Ugolini
*University of Toulouse, LEREPS*

8 October 2020*

## Abstract

This paper presents a detailed analysis of how liquid money market instruments – sterling bills of exchange – were produced during the first globalisation. We rely on a unique data set that reports systematic information on all 23,493 bills re-discounted by the Bank of England in the year 1906. Using descriptive statistics and network analysis, we reconstruct the complete network of linkages between agents involved in the origination and distribution of these bills. Our analysis reveals the truly global dimension of the London bill market before the First World War and underscores the crucial role played by London intermediaries (acceptors and discounters) in overcoming information asymmetries between borrowers and lenders on this market. The complex industrial organisation of the London money market ensured that risky private debts could be transformed into extremely liquid and safe monetary instruments traded throughout the global financial system.

# The Origination and Distribution of Money Market Instruments: Sterling Bills of Exchange during the First Globalisation

At the beginning of the twentieth century, the pound sterling's dominance over the global financial system was at its height. The City of London hosted the world's most important centre for international short-term lending and borrowing[1] and the main instrument for these money market transactions was the *sterling bill of exchange* (or bill on London). The sterling bill of exchange was a tradable asset originated by private agents located anywhere in the world to obtain short-term credit from London. It was one of the most liquid financial assets of the time and was traded in all significant financial centres.[2] Just before the First World War, about half of world trade was financed through this instrument.[3] Contemporaries generally considered that the sterling bill was 'a kind of world currency', 'the same as gold', or 'the equivalent of a bullion certificate'.[4] Many UK and foreign financial institutions were interconnected through the London bill market, whose importance for the international banking system was systemic.[5] The sterling bill was the cornerstone of the global financial system and its undisputed liquidity was key to the UK's financial dominance during the first age of globalisation.[6]

Why was the sterling bill of exchange considered such a liquid and safe credit instrument? And why did the London money market prove so robust in the decades prior to the First World War? While these questions have been asked before, any attempt to answer them has been confronted with a lack of systematic data on bill trading. Since the bill market was of the over-the-counter type, there was no central authority that recorded information on all London-originated bills. In this paper, we overcome this difficulty by relying on a detailed data set constructed from a unique archival source: the Bank of England's *Discount Ledgers*. These ledgers report systematic micro-level information on a profusion of bills circulating on the London money market and on all agents involved in their origination and distribution. Our data set contains information on all individual bills re-discounted by the Bank of England during the year 1906 (23,493 bills). Bills re-discounted by the Bank of England only represented a small share of all bills issued on the London market and did not constitute a fully representative sample. Nevertheless, the information they contain provides invaluable insights into the microstructure of the money market. We use descriptive statistics and network analysis to reconstruct the complete network of linkages between agents involved in the design of these bills. Doing so allows us to describe the industrial organisation behind the production of sterling bills in the early twentieth century.

---

[1] Keynes, *Treatise on Money,* p. 282.
[2] Jacobs, *Bank Acceptances*; Warburg, *Discount System*; Withers, *Meaning of money*; Flandreau and Jobst, 'Network Analysis.'
[3] Kynaston, *City of London*, p. 8; Atkin, *Foreign Exchange Market*, p. 5.
[4] Baster, *International Banks*, p. 13; Gillett Brothers, *Bill on London*, p. 16; Greengrass, *Discount Market*, p. 37.
[5] Bills on London formed part of the liquid reserves of UK commercial banks and of the foreign currency reserves of commercial and central banks abroad (Eichengreen and Flandreau, 'Central Banks').
[6] King, *Discount Market*, p. xi; Atkin, *Foreign Exchange Market*, p. 5.

Bills of exchange always involved a *drawer* (a borrower located either in the UK or abroad), an *acceptor* (a London-based actor which guaranteed the bill's payment in pounds sterling at maturity), and a *discounter* (the buyer of the bill). The data that we have assembled enable identification of how these various agents interacted on the money market.

[[ INSERT **Figure 1** about here ]]

Our analysis based on a new dataset confirms the true global dimension of the London bill market at the start of the twentieth century. We show that drawers were numerous and scattered across the world (Figure 1). Our data set records 3,554 different drawers, most of which were small private firms or merchants. Since the average investor could not hold detailed information about all these debtors, large information asymmetries must have existed between borrowers and lenders on the money market. Such information asymmetries constitute market frictions, which typically result in adverse selection and a total absence of lending.[7] Data on the intermediaries involved in the origination and distribution of bills allow us to demonstrate how these market frictions were overcome.

---

[7] Stiglitz and Weiss, 'Credit rationing'.

We argue that the information problem inherent to the production of sterling bills was solved by a cumulative process in which London intermediaries successively added their guarantees to the instrument. Acceptors played the key role in producing information on borrowers and in reducing credit risk. However, their signature was also supported by that of the discounters who purchased the bills before distributing them to other investors. When they resold (endorsed) a bill of exchange, discounters added their personal, secondary guarantee to it and thus enhanced its credit – a circumstance known as the *joint liability rule*.

Our results therefore shed light on the complex structure of the London money market at the start of the twentieth century and on the various mechanisms through which the information problem inherent to the production of bills was solved. These mechanisms allowed borrowers from around the world to access London credit facilities and ensured that risky, private debts could be transformed into the highly liquid and safe monetary instruments that were traded throughout the global financial system.

Our paper contributes to the historiography on the use of bills of exchange as instruments of international credit and payment and on the evolution of the London money market.[8] We integrate perspectives from the literatures on the organisation of the London discount market,[9] the business and operations of merchant banks,[10] and the role of joint liability in lending.[11] We show how these various factors jointly contributed to the high liquidity of the London money market.

The rest of our paper proceeds as follows. Section I explains how the bill of exchange functioned. In Section II we discuss our primary source and data. Section III underscores the role of acceptors in reducing information asymmetries on the bill market, while Section IV focuses on the role of discounters. Section V concludes.

---

[8] De Roover, *Lettre de change*; Scammell, *London Discount Market*; Nishimura, *Inland Bills*, Michie, *British Banking*; Cassis, *Capitals of Capital*; Mollan and Michie, 'City of London'.
[9] King, *Discount Market*; Scammell, *London Discount Market*.
[10] Chapman, *Merchant Banking*; Roberts, *Schröders*; Wake, *Kleinwort*.
[11] Ghatak and Guinanne, 'Joint Liability'; Santarosa, 'Long-Distance Trade'.

# I

From the late sixteenth until the early twentieth century, a negotiable bill of exchange was the standard financial instrument for obtaining short-term credit and exchanging currencies.[12] The bill of exchange was defined legally in the UK as *'an unconditional order in writing, addressed by one person to another, signed by the person giving it, requiring the person to whom it is addressed to pay on demand or at a fixed or determinable future time a sum certain in money to or to the order of a specified person, or to bearer.'*[13] Whereas 'sight bills' were payable after a few days, 'long bills' had a longer maturity (typically three or six months) and therefore served as short-term credit instruments.

A bill always involved at least three agents: a 'drawer', an 'acceptor', and a 'discounter'. The drawer was the person who addressed the bill; the acceptor, the individual or institution to whom the bill was addressed; and the discounter, the bill's beneficiary. By *accepting* the bill, the acceptor committed to pay the specified sum to the discounter at the specified date. Because the bill of exchange was a negotiable instrument, a discounter's claim on the acceptor could always be transferred to another investor (or *re-discounter*) at any time before maturity.

Although its legal form remained practically unchanged for centuries, the bill of exchange proved to be a flexible instrument that could be employed to finance diverse types of operations. Bills could be drawn to finance domestic and international trade transactions, to raise short-term money or to engage in purely financial operations such as security investment, currency speculation, or interest arbitrage.[14] Bills of exchange were traded in all principal cities in the world. However, bills accepted in London largely outnumbered those drawn from London on other foreign centres.[15] This reflected the importance of the sterling bill in the global financial system; it not only served to finance most of the UK's trade with foreign countries, but was also used to fund various commercial and non-commercial transactions taking place anywhere in the world.[16]

---

[12] De Roover, *Lettre de change*; Accominotti and Ugolini, 'International Trade Finance'.
[13] Article 3 of the 1882 Bill of Exchange Act.
[14] Goschen, *Theory of the foreign exchange*, pp. 23-42; Clare, *A B C of foreign exchanges*, pp. 80-87; Herger, 'Interest parity-conditions'.

Figure 2 presents an illustrative example of a transaction commonly financed through sterling bills in the early twentieth century: an export of goods from 'city A' to 'city B'. The figure's panel A shows the operations involved when the bill is issued. An exporter in city A has agreed to sell goods to an importer in city B (path 1 in the figure) but needs credit in order to finance production and shipment before receiving payment. The exporter (here, the *drawer*) draws a bill on a London agent (the *acceptor*) and asks for an engagement to pay to the bearer of the bill, at a specified date in the future, a sum in pounds sterling corresponding to the proceeds of the sale (path 2).[17] The drawer then transfers the bill (3) to her local bank (the *remitter*), which arranges to send it to a *discounter* in London (4). The discounter might have been pre-selected – either directly by the drawer (if she has London correspondents other than the acceptor) or by the acceptor herself (if she is the drawer's only correspondent in London).

[[ INSERT **Figure 2** about here ]]

Upon arrival of the bill in London, the discounter presents it to the acceptor (5); she 'accepts' the bill by countersigning it, thereby confirming her commitment to pay the bearer at maturity (6). Once the bill is accepted, the discounter credits the remitter's account (7). The remitter, in turn, credits the drawer's account (8) and so provides the financial means for producing and shipping the goods. The discounter can then either keep the bill until its maturity or resell ('endorse') it to a final investor (the endorsee, or *re-discounter*) willing to lend capital until the pre-specified date (9 and 10). The bill can be re-discounted an unlimited number of times before its maturity.

---

[15] Clare, *A B C of foreign exchanges*, p. 11.
[16] Ibid., pp. 11-15; Goschen, *Theory of foreign exchanges*, pp. 32-33.
[17] In this example, the bill is 'placed to the importer's account', which means that the importer authorises her exporter to draw on the acceptor with whom she is in a business relationship. However, if the drawer were in a direct business relationship with the acceptor then it would be said that the bill was 'placed to the drawer's account'.

Panel B of Figure 2 summarises the operations taking place when the bill expires. Just before maturity, the importer – who has, by then, received delivery of the shipped goods – remits funds directly to the acceptor (11); those funds enable the acceptor to meet the bill's payment. On the bill's actual maturity date, the bearer presents it to the acceptor for payment (12 and 13). Thus the instrument disappears at maturity, or 'self-liquidates' in the wording used by contemporaries.[18]

It should be clear from this example that a bill's acceptor did not advance her own capital; rather, she committed only to repaying the bearer in the expectation of receiving a monetary flow from the importer before maturity.[19] In other words, the acceptor was just a *guarantor* of the bill who added her signature to it – usually in exchange for a fee.[20] In contrast, the discounter and re-discounter immobilised their own funds in order to purchase the bill. These actors were (respectively) the first and ultimate lender. The usual procedure was for investors (re-discounters) in sterling bills to purchase them from a limited set of London institutions (discounters), who in turn had obtained those bills either from their correspondents abroad (remitters) or from acceptors. These first discounters constituted the 'wholesale' segment of the London discount market.[21]

Recall that every seller of a bill of exchange also had to 'endorse' it, thereby adding a secondary guarantee to the bill. In case the acceptor failed to pay the bill at maturity, the last endorser was liable for repaying the sum due to the bearer. By originating a bill, the drawer was thus able to borrow from an unknown lender (the re-discounter) thanks to the guarantee provided by an acceptor and to the intermediation – and secondary guarantee – of a wholesale discounter. The bill of exchange was not collateralised by any financial asset or 'physical' goods; it was secured instead via the guarantees provided by the successive intermediaries involved in its origination and distribution.[22] These intermediaries certified the quality of the bill's underlying debt and its repayment upon its maturity.

---

[18] Jobst and Ugolini, 'Coevolution', pp. 162-3.
[19] Nevertheless, acceptance houses reportedly followed certain rules in order to manage the risks of guaranteeing bills. One often cited rule was that they should not accept bills for more than three or four times the value of their paid-up capital and reserves (See Committee on Finance and Industry, *Minutes of Evidence*, vol.1, p. 73, par. 1204).
[20] Hawtrey, *Currency and Credit*, p. 129.
[21] Specialised wholesale *discount brokers* (connecting first discounters to re-discounters) had already emerged in London by the end of the nineteenth century. For instance, Sayers, *Gilletts*, pp. 51–52 writes that the discount house Gillett Brothers & Co. (a leading London re-discounter) used to purchase, in the 1890s, its entire portfolio of Indian-drawn bills through the intermediation of discount brokers Page & Gwyther.

The specific type of transaction described in Figure 2 was common in the early twentieth century; yet bills of exchange could be mobilised in many other ways, and all these roads led to the London money market. A detailed exposition of the manifold uses of bills can be found in a handbook published by one of London's foremost discount houses.[23] Bills were first used to finance trade. Especially in early times, the drawer was often the seller of some goods and the acceptor their buyer. By signing the bill, the acceptor promised to pay the value of the sold goods after their delivery – thus allowing the seller to raise capital and finance shipment. In that case, the bill's acceptor was a UK importer.[24] Bills drawn directly on importers were called *trade bills* on the London discount market.

From the mid-nineteenth century onward, several trading and financial houses in the City began offering their respective signatures and allowed exporters to draw bills upon them rather than on their importers.[25] Bills accepted by reputable financial institutions were known as *bank bills* and were usually considered superior to trade bills because of the acceptor's higher standing. Hence, contemporaries referred to such bills as 'first class paper' as opposed to the 'lower class paper' drawn on less reputable, non-financial firms. In these cases, the acceptor was not engaged in the commercial transaction; instead it was a third party that agreed to accept bills in the importer's name – on the condition that the latter (privately) agree to provide the funds needed to meet the bills' payment at maturity.[26] Bills could also be drawn directly by the importer (rather than the exporter) on the financial house with whom she had the arrangement. In this case, the importer raised capital herself to finance the goods' shipment.[27] Sometimes, the acceptor did not have a direct relationship with the drawer but only with her bank, which took care of selling the bill to a discounter and of providing the funds to the acceptor before maturity. In such cases, often the drawer's bank also endorsed the bill before it was accepted.[28]

---

[22] Flandreau and Ugolini, 'Lending of Last Resort'. The holder of an unpaid bill could not seize the commodities that it financed; the only recourse was to seize the acceptor's or previous endorser's assets.
[23] Gillett Brothers, *Bill on London*.
[24] Ibid., pp. 47-48
[25] Greengrass, *Discount Market*, p.46.
[26] Hawtrey, *Currency and Credit*, pp. 123-24; Gillett Brothers, *Bill on London*, pp. 27-29, 37-39, 41-43. In contemporary parlance, the importer engaged to 'make provision' and to 'cover' the acceptor before maturity. This particular case corresponds to the example described in Figure 2.

As stated above, bills could also be used to finance activities other than trade. For instance, the drawer might be an industrial firm that needed a short-term, blanked credit to finance production and sought to raise capital on the London discount market with the guarantee of an acceptor.[29] Should the firm's production remain unachieved at the bill's maturity, then the acceptor could authorise the drawer to draw another bill so that the debt could be rolled over.[30] Finally, in many cases the drawer was a financial firm just willing to fund its own stock or bond investments or to refinance its banking operations. By originating a bill, a financial firm could replenish its liquidity while using the acceptor's guarantee as collateral.[31] Bills originated for purposes other than trade were referred to as *accommodation* or *finance bills*. Although the practice of drawing bills that were not based on 'genuine' commercial transactions was often decried, the distinction between 'finance' and 'real' bills was not clearly defined. For example, contemporaries debated whether bills drawn to finance a future, expected commercial transaction should be classified as accommodation paper.[32] Although certain bills provided details about the nature of the commercial transaction they financed (e.g, details on the sold goods and their shipment), doing so was optional (and irrelevant from the judicial standpoint) and in most cases investors could not directly recognise a 'finance' bill from a 'real' one. Hence, the standing of bills on the London discount market did not mostly depend on their intrinsic nature but on the reputation of the intermediaries which had guaranteed them.[33]

---

[27] Gillett Brothers, *Bill on London*, pp. 29-31, 39-40.
[28] Ibid., pp. 53-55.
[29] Goschen, *Theory of the foreign exchanges*, pp. 38-41.
[30] Gillett Brothers, *Bill on London*, pp. 45-47.
[31] Also in this case, the acceptor was not necessarily in a direct customer relationship with the drawer; thus, for example, the former might only have had an arrangement with the drawer's correspondent.
[32] Goschen, *Theory of the foreign exchanges*, pp. 38-41.

The information recorded on a bill of exchange allowed bearers to reconstruct many, but not all, of the underlying interlinkages that supported its origination and distribution. Simply looking at a bill was not enough to reliably determine the exact nature of the transaction that stood behind it. That said, each bill did record the name of its drawer as well as the names of all intermediaries who had guaranteed and/or purchased it.

[[ INSERT **Figure 3** about here ]]

Figure 3 transcribes a typical bill of exchange found in the archives of the leading acceptance house Kleinworts & Co. This £3,000 bill was drawn on 10 August 1910 by the Moscow-based Société L. Bauer & Co. (the drawer) and was made payable after three months by Kleinworts & Co. (the acceptor). After drawing the bill, the drawer immediately sold it to the Banque de Commerce de l'Azow-Don/Azow-Don Commerzbank (the remitter), probably the drawer's bank in Moscow. The very same day (10 August), the remitter sold/endorsed the bill to the Union Discount Company of London (the discounter), which thus became entitled to cash it in at maturity. Azow-Don Commerzbank shipped the bill to London, where Kleinworts accepted it (by affixing its signature) on 15 August before transmitting it to the Union Discount Company. The discounter kept the bill until maturity and did not resell it to another investor. Three months later, Kleinworts therefore repaid the Union Discount Company £3,000 through a London clearing bank (the London County & Westminster Bank), which was responsible for pure payment services. The discounter then returned the self-liquidated instrument to the acceptor, in whose archives it remains preserved.

The information recorded on that bill does not indicate the exact nature of the transaction it financed. The drawer (Bauer & Cie) might have been an exporter of Russian goods. Yet because the bill makes no mention of any shipment of goods, we cannot be sure that it was used to finance trade. Inspecting the bill itself leaves us in the dark also with regard to the exact nature of the relationships between the various parties involved. For example, the remitter (Azow-Don Commerzbank) might have selected the discounter (the Union Discount Company) directly; alternatively, the acceptor (Kleinworts) might have arranged for the bill to be discounted. In that event, Kleinworts would have instructed the Moscow bank to endorse the bill to the Union Discount Company (the discounter) before shipping it to London.[34]

---

[33] Gillett Brothers, *Bill on London*, p. 22.

One must bear in mind that, even if all details of the transaction were not known, a bill's purchaser could always identify the most important actors involved in its production. In particular, a bill recorded the names of all intermediaries whose signatures amounted to collateral for it. Those agents included the drawer (or borrower; here, Bauer & Cie); the acceptor (or guarantor; here, Kleinworts & Co.), and the discounter (or lender; here, the Union Discount Company).

The largest acceptors of bills in London were the merchant banks or acceptance houses that specialised in offering acceptance services for their customers at home and abroad.[35] Acceptors also included UK deposit banks, branches of foreign banks, and 'Anglo-foreign banks' – multinational banks based in London but whose business was concentrated in certain foreign geographical areas, where these banks specialised and maintained a large network of correspondents.[36] In addition, a large number of UK trading or manufacturing firms also accepted bills drawn on them by their trading partners.[37]

Among the largest discounters were the so-called discount houses of the City. These highly specialised institutions purchased large amounts of bills, which they then kept in their own portfolios or re-discounted to other investors.[38] Discount houses usually funded their investments with short-term deposits or 'call money' from other financial institutions (especially the large UK deposit banks).[39] However, discount houses were not the sole distributors of bills on the money market. Foreign and Anglo-foreign banks also played that role, while trading and manufacturing firms discounted bills as well. In contrast, UK deposit banks invested in (i.e., re-discounted) bills but seldom served as wholesale sellers on the discount market.[40]

---

[34] The Union Discount Company was a 'discount house' that was not actually involved in the business of correspondent banking. Thus, it is unlikely that this company was the London correspondent of Azow-Don Commerzbank. Hence we suspect that Kleinworts both accepted the bill and found a discounter (the Union Discount Company) willing to purchase it in London.
[35] Greengrass, *Discount Market*; Chapman, *Merchant Banking*.
[36] Jones, *British Multinational Banking*. Anglo-foreign banks are also often referred to as British overseas banks or British multinational banks.
[37] Sayers, *Gilletts*.

## II

Through its monetary operations, the Bank of England was an influential player in the London discount market. Large holders of bills approached the Bank – in times of monetary tension and before publication of their balance sheets – for re-discounting and thus obtaining cash.[41] The Bank of England gathered systematic information on all the bills it re-discounted, thereby monitoring its exposure to acceptors and discounters.[42] In its *Discount Ledgers,* it recorded complete information on the identity of the intermediaries (drawer, acceptor, discounter) involved in the origination and distribution of re-discounted bills.[43]

We collect information on all bills re-discounted by the Bank of England during one year. We select a period equal to exactly one year in order to circumvent any seasonality concerns. Because the Bank only acquired bills through standing facility lending and never through open market operations, its bill portfolio only became sizeable in times of monetary tension.[44] To ensure our dataset captures a significant portion of the sterling bill market, it is important to select a year in which the Bank's discount window was rather active. The year 1906 is a good candidate here as the UK's external position varied throughout the year leading the Bank of England to alter its discount rate on six different occasions even though there was no full-blown financial crisis as in 1890 or 1907.[45] Figure 4 shows the evolution in the number of bills re-discounted by the Bank of England from 1889 to 1910, as well as the share of these re-discounted bills in the total amount of sterling bills issued on the market as estimated by Nishimura.[46] The figure shows that in 1906, the Bank re-discounted a higher share of sterling bills issued than in any other year, although that proportion remained limited overall (2.47 per cent). Bills re-discounted by the Bank in 1906 had an average value of £1,346 (compared to £1,556 on average in 1900-1910), an average maturity of 44 days (compared to 46 days), and were purchased at the average interest rate of 4.17 per cent (3.45 per cent).

---

[38] Vigreux, *Crédit par Acceptation*, pp. 169-70; Sayers, *Gilletts*, pp. 37-38.
[39] For a detailed description of the business of discount houses, see King, *Discount Market*, Scammell, *London Discount Market*, Fletcher, *Discount Houses*, and Cleaver, *Union Discount*.
[40] Spalding, *Foreign exchange*, p. 200; Hawtrey, *Currency and Credit*, pp. 130.
[41] Tensions on the money market could occur when the demand for cash was unusually high, thus putting upward pressure on market interest rates. Such circumstances could arise from concerns about the UK's external balance or about the position of certain financial institutions, as well as from seasonal liquidity demands. UK financial institutions also re-discounted a large portion of their bills to the Bank of England just before publication of their balance sheet in order to increase the published amount of their cash reserves. This practice was known as *window-dressing.*
[42] Flandreau and Ugolini, 'Lending of Last Resort'
[43] Bank of England Archives, *Discounters' Ledgers* (C22/46-50), *Drawing Office Discounters' Ledgers* (C23/7), *Bankers' Ledgers* (C24/6), *Upon Ledgers* (C26/72-74).
[44] Ugolini, 'Liquidity Management'.

[[ INSERT **Figure 4** about here ]]

The *Discount Ledgers* contain the accounts of the Bank's clients. During any re-discounting operation, the Bank registered the bill's information in the discounter's account (in a column labelled 'with') and also in the acceptor's account (in a column labelled 'upon').[47] Thus each bill re-discounted by the Bank was recorded twice in the *Ledgers*. To avoid recording the bills twice in our database, we collect only the 'upon' entries of the Bank's *Ledgers*.[48] For each bill, we record the name and location of the three parties involved in its origination and distribution: the drawer, the acceptor, and the discounter.

---

[45] Flandreau and Ugolini, 'Lending of Last Resort' show that the quality of bills re-discounted by the Bank of England was not altered in crisis times. Nevertheless, bills rediscounted by the Bank during a full-blown financial crisis might not have been representative of the money market.
[46] Nishimura, *Inland Bills*.
[47] See Flandreau and Ugolini, 'Lending of Last Resort' for a description of the Bank's accountability in the re-discounting of bills.
[48] We prefer to collect the 'upon' entries because the 'with' entries are sometimes less detailed. Bills were usually re-discounted not individually but rather in packs known as 'parcels'. A specific category of the Bank of England's *Discount Ledgers* includes the accounts of discount houses (*Brokers' Ledgers*, C25/5). In these ledgers, the parcels of bills discounted were not always 'unpacked' in the 'with' entries. Yet in the 'upon' entries the parcels *were* unpacked – that is, under the headings of the acceptors of each bill contained in the parcel.

We use these data to describe relations between agents on the London money market. From our data set for 1906, which contains 23,493 bills, we reconstruct the complete network of agents whose names appear on the bills. In this way we obtain a static network of 4,970 agents, or 'nodes'. Among these we find that the drawer role is played by 3,554 nodes, the acceptor role by 1,439, and the discounter by 145 nodes (note that some nodes played more than one role).

We record all relationships, or *links*, between triplets of agents in the network. We define two direct relationships between pairs of actors: between drawers and acceptors, and between acceptors and discounters. Thus a link exists between a given drawer and a given acceptor when the latter has accepted at least one bill drawn by the former, and there is a link between an acceptor and a discounter when the latter has discounted at least one bill accepted by the former. As a result, there also exists an indirect relationship between a drawer and a discounter (through an acceptor) when the latter has discounted at least one bill drawn by the former.

There are, of course, some limitations to our source and resulting data set. As far as we know, the Bank of England's *Ledgers* constitute the most comprehensive and detailed source on the sterling bill market. Nevertheless, bills re-discounted by the Bank only represented a small portion of all sterling bills issued, and this sample might not have been representative of the entire market.

For one thing, the Bank of England only re-discounted bills from a limited set of discounters that it declared 'eligible'. Eligible discounters were not representative of the final investor in sterling bills (the re-discounter in Figure 2). Our archival source does not allow identifying whether the agents who sold bills to the Bank of England were those bills' first discounters or if they had themselves bought the bills from other investors. However, eligible discounters made up the wholesale segment of the London discount market (the discounter in Figure 2), and therefore included all institutions that purchased bills through acceptors or foreign correspondents and then resold them to final investors (or to the Bank of England). This group of institutions included discount houses, Anglo-foreign and foreign banks, and merchant banks as well as non-financial, trading firms.

Biases might also have existed in the nature and quality of bills re-discounted by the Bank as compared with bills sold on the open market. Since there was no formal rule regarding the eligibility of drawers or acceptors, the Bank of England could in theory re-discount bills drawn or accepted by all sorts of agents.[49] However, it is possible that the Bank was a particularly cautious re-discounter, in which case one would expect the bills it purchased to be, on average, of higher quality than those circulating on the market. Alternatively, it is also possible that eligible discounters exploited the Bank's discount window strategically and re-discounted their lowest-quality bills in order to keep the best ones on their own balance sheet. If that was the case, one would expect the Bank's bill portfolio to disproportionately consist of bills accepted by weaker institutions.[50] Although a vast literature has discussed the Bank of England's rediscount policy, no attempt has been made so far to empirically assess the extent and direction of biases in its bill portfolio.

We perform three series of cross-checks on our data. First, Jansson reports the aggregate amounts of bills accepted by nine top acceptance houses.[51] We can rank these nine houses according to their total amount of outstanding accepted bills at the end of 1906, and compare the ranking of the same institutions in the Bank of England's portfolio of re-discounted bills that same year (table 1).[52] The result is reassuring, as the two rankings almost perfectly match. The ranking of acceptors also exhibited little year-on-year variation.[53]

---

[49] There was no constraint on drawers, while the only constraint on acceptors was that they had to be based in London in order to allow for the collection of bills' repayment at maturity. The bills accepted by intermediaries who were not among the Bank's agreed customers were those recorded in the so-called *Upon Ledgers*. In 1906, they made for 23.39 per cent of all bills re-discounted.
[50] Here, we use the term 'portfolio' to denominate the entire set of bills acquired by the Bank of England throughout the year 1906 (rather than the Bank's bill holdings at one point in time).
[51] Jansson, *Finance-Growth Nexus*.
[52] The ranking of the selected houses in the Bank of England's portfolio is based on the number of discounters that purchased bills accepted by them.
[53] Jansson, *Finance-Growth Nexus*, p. 209

[[ INSERT **Table 1** about here ]]

Second, we were able to retrieve from the archives of two major acceptance houses (Kleinworts and Barings) the list of their top 10 clients in 1906.[54] We check whether these clients' names also appear as drawers of bills accepted by Kleinworts and Barings and re-discounted by the Bank of England. Six (respectively, five) out of Kleinworts' (respectively, Barings') top 10 drawers also appear on bills in our dataset. This suggests that at least the largest drawers of sterling bills were well represented in the Bank's portfolio.

[[ INSERT **Table 2** about here ]]

Finally, the archives of one major discount house (Gillett Bros. & Co.) contain a detailed breakdown of the bills it purchased in 1906 by each principal acceptor.[55] We can thus compare the entire portfolio of bills which Gilletts discounted on the market with the small portion of that portfolio (0.85 per cent) which the firm re-discounted to the Bank of England. Table 2 shows how different categories of acceptors feature in each sample. The two samples are broadly consistent, even though we do notice discrepancies. In particular, bills accepted by Anglo-foreign banks are clearly over-represented in the sample of bills re-discounted to the Bank. Gilletts also discounted a significant amount of bills accepted by foreign banks' London branches (13.1 per cent of its total discounts) but did not re-discount any of those bills to the Bank of England. At the same time, there is no evidence of a quality bias in Gilletts' re-discounts to the Bank of England. Bills accepted by non-financial firms (trade bills) figure in similar proportions in both samples. While merchant banks are slightly under-represented in Gilletts' rediscounts to the Bank, we do not find any systematic bias in favour of the most reputable ones. Several first class signatures such as

---

[54] The list of Kleinworts' top-10 clients was constructed from London Metropolitan Archives, CLC/B/140/KS04/12. The list of Barings' top-10 clients was constructed from the Baring Archive, 202368-202376, *Credit Department Annual Report 1906-1914* (file communicated by the archivists).

[55] London Metropolitan Archives, CLC/B/100/MS24688/002. Note that this ledger only covers the 135 top acceptors that Gilletts held in their portfolio. Sayers, *Gilletts*, p. 46 writes that in 1905, these top acceptors accounted for 73 per cent of the house's total discounts. Jansson, *Finance-Growth Nexus*, pp. 252-5, relies on the same archival source to describe the evolution in Gilletts' bill portfolio over 1892-1913, and shows there was little volatility in the acceptor composition of the portfolio.

Rothschilds and Brown Shipley are slightly over-represented at the Bank, but other equally reputable institutions such as Kleinworts and Schroders are slightly under-represented (Figure 5).

[[ INSERT **Figure 5** about here ]]

In addition, we also compare acceptors in Gilletts' bill portfolio and in the Bank's entire bill portfolio (across all discounters). Out of the 127 acceptors appearing on bills discounted by Gilletts in 1906, only 11 are absent from our dataset. These include 6 foreign banks, 1 Anglo-foreign bank and 4 non-financial firms. Table 3 displays the correlation between the importance of the various acceptors in Gilletts' bill portfolio and in our dataset. The coefficient of correlation is 0.52 and goes as high as 0.62 when we exclude foreign banks, which are under-represented at the Bank. The correlation is also strongly positive and statistically significant across all categories of acceptors.

[[ INSERT **Table 3** about here ]]

Existing sources do not allow systematically comparing bills re-discounted by the Bank of England with the whole population of sterling bills circulating on the London market. It is evident that bills re-discounted by the Bank only represented a small share of the bill market and were not randomly selected. However, our comparison of the Bank of England's bill portfolio with that of one significant private actor on the money market does not reveal manifest quality differences between the two institutions.[56] We do however identify an interesting pattern. Compared to the rest of the discount market, the Bank re-discounted more bills accepted by Anglo-foreign banks and fewer bills accepted by foreign banks. Contemporaries were aware of the Bank's reluctance to re-discount bills drawn on foreign banks. One source for example described how bills accepted by certain foreign banks were considered 'first class' in the market but were 'tabooed' by the Bank of England.[57] This policy may have aimed at supporting British financial institutions in their competition with foreign banks in the acceptance business.

---

[56] We cannot conclude through a simple cross-check of our data with external sources whether biases in the Bank of England's portfolio of re-discounted bills were due to active discrimination on the Bank's side or to the discounters' own choice to disproportionately re-discount certain types of bills.

# III

We first use our data set to document where the debts underlying sterling bills re-discounted by the Bank of England were originated. Figure 1 showed how the drawers of bills were dispersed geographically. Among all the drawers of bills re-discounted in 1906, UK drawers represented only 13.56 per cent; 17.50 per cent of drawers were located in continental Europe, 20.40 and 15.14 per cent were in (respectively) USA/Canada and Latin America, 19.78 per cent in India and the Far East, 5.46 per cent in Africa, 2.11 per cent in Oceania, and 6.05 per cent in the rest of the world.

Not all the drawers of sterling bills were located in the world's largest metropolises or trading centres; many originated from cities with much smaller populations. This phenomenon is evident from the geographical location of European drawers, which Figure 6 shows were scattered across the continent. Many drawers of bills were located in smaller localities – especially in Central Europe, Scandinavia, Spain, and Italy – from which we conclude that many foreign local firms had access to London credit facilities. Thus it appears that, at the beginning of the twentieth century, firms from all around the world could borrow on the London bill market.

[[ INSERT **Figure 6** about here ]]

How could such diverse and geographically widespread borrowers gain access to the London money market and borrow through sterling bills? The investor in bills of exchange could barely rely on hard indicators to assess the borrowing firms' solvency (let alone their honesty), and their geographical dispersion made it difficult for a distant investor to assess conditions in the various markets where they operated. It follows that there must have been severe information asymmetries between borrowing firms and final investors on the London money market. Such market frictions could well have resulted in credit rationing for borrowers and an absence of lending.[58]

---

[57] Kynaston, *City of London*, p. 282.

In order to understand how these frictions were overcome, it is essential to look at the role of intermediaries in the production of sterling bills. Before it reached the final investor, each bill was first accepted/guaranteed by an acceptor and subsequently distributed by a discounter. As we have explained, London acceptors were the guarantors of sterling bills. In case the drawer (or her trading partner) failed to reimburse her debt, the bill remained the acceptor's liability: she was still obliged to repay its bearer at maturity. An acceptor was the first exposed when borrowers defaulted, so she was strongly incentivised to gather detailed information about them.

Among the largest acceptors in the City were the specialised acceptance houses, which accepted bills drawn by their numerous domestic and foreign clients. Archival records of the merchant bank Kleinwort & Co. illustrate the role of these houses in producing private information about borrowers seeking to access the London money market. Founded in 1855, Kleinwort & Co. gradually established itself as a major acceptance house over the second half of the nineteenth century; by 1906, it was the leading acceptor of sterling bills (see Table 1). The firm typically offered credit lines under specific conditions to its customers around the globe. Under these arrangements, Kleinwort & Co. committed to accept bills (up to a certain amount) on account of its customers. The conditions of the credit lines – in particular, their total amount and the commission charged for accepting bills – varied as a function of the borrowing firm's credit standing.[59] In order to obtain information on its clients abroad, the house relied on its large network of foreign correspondents to produce detailed reports on clients' positions that described these firms' origins and commercial activities while assessing their financial situation (especially their capital) and the owners' personal qualities. These reports, which were often written in a foreign language, were gathered into 'client information books' and updated frequently.

---

[58] Stiglitz and Weiss, 'Credit rationing'.
[59] See, for example, the *Client Account Ledgers* at London Metropolitan Archives, CLC/B/140/KS04/12/22.

The type of information gathered about borrowers could be acquired only through frequent contacts with those clients and was rarely quantifiable; thus it was 'soft' information.[60] For example, the information book on German customers described Heine & Fleich – a family business, located in Altona (Hamburg), that specialised in the trade of leather, skins, and furs – as a 'reputable firm' whose 'financial situation is favourable' and is 'considered solvent for its orders'. This report added that, 'on a personal note, the owners are described to us as competent and respectable merchants'. The case of Kleinworts therefore suggests that acceptance houses acted as relationship bankers toward their clients who wanted to borrow on the London money market. Through repeated interactions with these clients, acceptance houses gathered private information about bill market borrowers.[61]

We hypothesise that, through their information acquisition activities, acceptors made an indispensable contribution to resolving information asymmetries between borrowers and lenders on the money market. To provide quantitative evidence for this proposed mechanism, we rely on the theory of relationship banking. According to this literature, firms on which little public information is available usually borrow from only one or a small number of creditors.[62] Private information about borrowers can be acquired only through repeated transactions, and there are fixed costs involved. Therefore, lending to such borrowers is more efficiently handled by one single intermediary (or a small number of them). In contrast, firms whose standing and creditworthiness are publicly known will more efficiently borrow from a large set of creditors or directly from the capital market.[63] If the activity of accepting (guaranteeing) bills required private information about drawers (borrowers), then we should similarly expect the latter's bills to have been guaranteed by a small number of acceptors. Yet if the acceptor's guarantee had solved the information problem on the bill market, then drawers should have been able to sell their accepted bills to a larger number of discounters. Hence we check for whether the drawers of sterling bills were, on average, connected to more discounters than acceptors.

---

[60] Stein, 'Information Production', p. 1892 defines *soft information* as 'information that cannot be directly verified by anyone other than the agent who produces it'. In contrast, *hard information* is 'verifiable information, such as the income shown on the borrower's last several tax returns'.

[61] See Boot, 'Relationship Banking', p. 10 for a definition of *relationship banking*. The fee charged by acceptance houses compensated them for these information acquisition activities. On the information role of acceptance houses, see Accominotti, 'London Merchant Banks' and ' 'International banking and transmission of the 1931 financial crisis' and Flandreau and Mesevage, 'The separation of information and lending'.

[62] Diamond, 'Financial intermediation'; Sharpe, 'Asymmetric Information'; Diamond 'Monitoring and reputation'; Rajan, 'Insiders and Outsiders'; Peterson and Rajan, 'Benefits of lending relationships'; Berger and Udell, 'Relationship banking and lines of credit'; Boot, 'Relationship banking'; Boot and Thakor, 'Can relationship banking survive competition ?'

Our empirical strategy consists of comparing how acceptors and discounters (two different categories of *principals*) established relationships (or *links*) with drawers (or *agents*) on the bill market. We first focus on the 1,381 drawers whose names appear on at least two non-identical bills in our data set.[64] In Table 4, panel A reports the average number of acceptors and discounters per drawer. Although there are 1,439 different acceptors appearing in our data set and only 145 discounters, drawers of bills were on average connected to a smaller number of acceptors (2.83) than discounters (3.33). Whereas the ratio of the acceptor population to the discounter population is 9.92, the median acceptor-to-discounter ratio of drawers is only 1.16. As shown in panel B of the table (row 'All>1/Observed'), about half of the 1,381 drawers whose names appear on more than one bill had a strictly higher number of discounters than acceptors. In contrast, only 28.67 per cent of the drawers had more discounters than acceptors. This result holds irrespective of the number of transactions in which drawers were involved. Both small drawers (whose names appear on a limited number of bills) and large ones (that were involved in a much higher number of transactions) had, on average, fewer acceptors than discounters (see panel B, rows 'Observed').

[[ INSERT **Table 4** about here ]]

---

[63] Boot and Thakor, 'Can relationship banking survive competition?'
[64] Drawers for which only one transaction is recorded were, by construction, linked to just one acceptor and one discounter – which prevents us from drawing any conclusions about the structure of their personal linkages. Among the total of 3,554 drawers, 1,381 appear more than once in our data set.

Does this pattern reflect a true privileged relationship between drawers and acceptors, or does it arise from pure chance? In other words, given the demography of our network (the relative number of drawers, acceptors, and discounters), would we obtain the same result if links between agents had been formed purely randomly? In order to check this, we follow a standard methodology in social and banking network analysis which consists in comparing observed, 'real world' networks to random ones.[65] More precisely, we compare the distribution of drawers' acceptor-to-discounter ratio with that of a benchmark network (or 'null model') when link formation between drawers and acceptors/discounters is randomised. To that end, we perform two simulation exercises in which we generate 100 random networks (based on the Bernoulli graph model) with the same demography as the observed one but with simulated links (see Appendix A for details). In Simulation 1, random networks have the same number of nodes and the same number of links as the actual network, but we assume that links between nodes were formed in a purely random way – so that every acceptor/discounter had the same probability of forming a link with a given drawer. In Simulation 2, we again randomly recombine links between drawers and acceptors/discounters, but assume that each acceptor/discounter also maintained the same total number of links (with drawers) in the simulated network as in the actual one. Simulation 2 therefore allows assessing whether the pattern observed in the data does not arise from the individual characteristics of the nodes and, especially, the fact that there existed certain influential acceptors/discounters with a greater likelihood of establishing relationships with drawers. Panel B (rows 'Simulation 1' and 'Simulation 2') in Table 4 classifies drawers in the simulated networks according to whether they had more acceptors or discounters; Appendix A provides full details on simulated and observed distributions of the drawers' acceptor-to-discounter ratio. Appendix B provides summary statistics for various indicators describing the structural properties of the actual and simulated networks.

---

[65] See Wasserman and Faust, *Social network analysis*; Nier et al., 'Network models'; Iori et al., 'Italian overnight money market'; Chinazzi et al., 'International financial network'; Craig and von Peter, 'Interbank tiering'; Martinez-Jaramillo et al., 'Mexican banking system's network'.

If links between nodes had been formed in a random manner, then only a small minority of drawers would have had more discounters than acceptors (0.79 or 4.26 per cent, versus 50.25 per cent in the observed network). Some 40 per cent of the drawers in Table 4 appear on only two different bills. If these small, two-transaction drawers had chosen their acceptors/discounters randomly, then an overwhelming majority of them would have had a different acceptor and a different discounter for each bill – and therefore as many acceptors as discounters overall. In the actual network, however, only 26.70 per cent of the two-transaction drawers had as many acceptors as discounters whereas 47.67 per cent of them had two discounters but only one acceptor. When instead focusing on the largest drawers (those that had more than 10 transactions), we see that, in the observed network, 50.00 per cent of them had more discounters than acceptors. By contrast, if these big drawers had formed their relationships with acceptors and discounters randomly, only a small minority of them (3.24 or 5.71 per cent) would have had more discounters than acceptors whereas a majority (55.85 or 82.62 per cent) would have established links with a strictly higher number of acceptors than discounters. Thus the evidence indicates that drawers tended to maintain a few relationships with acceptors but had access to a larger pool of discounters. Most drawers of bills could deal only with the limited number of acceptors that held information on them.

Finally, we assess the extent to which acceptors had drawers in common – that is, to what extent they 'shared' drawers. Had acceptors held proprietary information about their drawers, it seems unlikely that drawers would be shared among acceptors. Our findings support this hypothesis. Acceptors tended to share very few drawers (and often, none) with other acceptors. In fact, 40 per cent of the acceptors in our data set did not share any of their drawers, and no acceptor shared drawers with more than 13 per cent of the other acceptors. Yet if the links between drawers and acceptors had formed randomly, then acceptors would be much more likely to share drawers (see Appendix A).

The acceptors' tendency not to share their drawers is characteristic of markets in which intermediaries hold proprietary information about their customers. Acceptors specialised in guaranteeing the debts of a few borrowers, on which they had acquired information and with whom they had special relationships. The evidence therefore suggests that acceptors played an important role in producing information about borrowers on the London money market and in overcoming market frictions.

## IV

We next explore how information problems shaped the market structure of the accepting and discounting industries in London. As we have described, the acceptor's guarantee was crucial in the investors' willingness to purchase bills on the money market. Of course, the value of this guarantee depended heavily on the acceptor's reputation. Reputational effects could have resulted in a high concentration of the accepting industry, since a few large acceptors might have been able to capture the reputational rents associated with guaranteeing commercial debts. Indeed, Chapman argues that, during the second half of the nineteenth century, the accepting business became increasingly concentrated around a few specialised merchant banks and acceptance houses.[66] At the same time, the acceptors' information acquisition activities might have suffered from diseconomies of scale. Small, decentralised institutions are widely considered to be more efficient (than are large, hierarchical ones) at acquiring and processing soft information about borrowers.[67] The reason is that the information derived by a bank officer is often difficult for upper management to verify. For example, the qualitative information that Kleinworts obtained about the owners of Heine & Fleich could hardly be verified by anyone other than the agent who had produced it. These diseconomies of scale in the acquisition of soft information could have constrained acceptors' capacity to grow.

---

[66] Chapman, *Merchant Banking*.
[67] Stein, 'Information Production'.

[[ INSERT **Table 5** about here ]]

Table 5 presents indicators of market concentration in the accepting and discounting industries, constructed from the sample of bills re-discounted by the Bank of England.[68] The table provides common measures of concentration: the *Herfindahl–Hirschman index* (HHI),[69] *highest market penetration*,[70] and the *market share*[71] of the top discounters and acceptors in our data set. According to this evidence, the accepting industry did not exhibit a high degree of market concentration. There is actually much greater concentration in discounting than in accepting: the HHI index is almost four times higher for discounters than for acceptors, and the greatest market penetration is twice for discounters what it is for acceptors. Similarly, the top 15 discounters in our data set captured more than 70 per cent of the market share in discounting, whereas the top 15 acceptors accounted for only a 35 per cent market share in their activity.[72] These results suggest that the very nature of accepting activities, which required maintaining personal relationships with customers abroad, resulted in diseconomies of scale and therefore limited market concentration in the industry.

[[ INSERT **Figure 7** about here ]]

---

[68] The construction of these indicators is based on the number of drawers per acceptor and per discounter.

[69] The Herfindahl–Hirschman index is defined as the sum of the squares of the market shares of all market participants. The index ranges from 0 (in the case of a perfectly competitive market) to 10,000 (in the case of a perfectly monopolistic market).

[70] A firm's market penetration is defined as the share of potential customers it reaches. Market 'penetration' differs from market 'share' in this sense: shares cannot be appropriated by more than one firm, but any number of firms can reach the same customer(s) at the same time. The highest market penetration is the penetration of the firm that reaches the largest number of potential customers. This metric can range from 0 per cent (when each firm reaches only an infinitesimal share of customers) to 100 per cent (when at least one firm manages to reach all potential customers).

[71] We compute the market share by treating each drawer–acceptor and drawer–discounter relationship as the unit portion of the existing market and then computing their sum. Thus we view the accepting (resp. discounting) market as consisting of 6,075 (resp. 6,758) portions.

[72] The lower market concentration in accepting is due to the fact that a myriad of small and large acceptors coexisted on the money market apart from the most significant acceptance houses (or merchant banks). Although these houses stood among the most significant acceptors in London, they did not enjoy a dominant position in this market.

In Figure 7 we report the frequency distribution of acceptors and discounters with regard to the number of drawers with whom they were connected in the set of re-discounted bills. Both industries were characterised by a 'dual' market structure in that many small actors co-existed with a small number of much larger ones. Most (64.42 per cent) of the acceptors in our data set were connected to one drawer only. These small acceptors were usually trading or manufacturing firms, not financial institutions. At the other end of the spectrum, a small minority (0.83 per cent) of acceptors were connected with more than a hundred drawers. These included the main commercial banks and London acceptance houses such as Kleinwort & Co. (the largest acceptor in our data set), which accepted bills drawn by 325 different drawers.

The discounting industry also exhibited a dual market structure. Among the discounters in our data set, 35 per cent of them were connected with only one drawer and 35 per cent were connected to more than 10 drawers. Yet discounters, unlike acceptors, seem not to have faced diseconomies of scale; those that managed to grow did so to a much greater extent than did acceptors. As a result, a small number of large discounters dominated the market, while a large number of small ones undertook much more limited discounting activities. These differences in market structure explain the higher level of concentration observed in the discounting than in the accepting industry. The largest discounters included the City's leading commercial and merchant banks as well as specialised discount houses such as the Union Discount Company (the largest discounter in our data set), which purchased bills drawn by 705 different drawers.

A consequence of the market structure just described – and of the limited market power of the large acceptance houses – was that a significant share of the bills produced in London were accepted by small, non-financial firms of modest reputation and on which little public information was available. How could bills drawn on such small acceptors end up on the money market and be brought to the final investors' portfolio?

We argue that discounters played an important role in reducing the risk inherent to these bills. After being accepted, sterling bills were purchased by a discounter who then distributed them to a final investor (a re-discounter). In this process, discounters endorsed the bills and added their personal, secondary guarantee to them – that is, in addition to that of the acceptor. Thus discounters served two distinct functions on the money market: they not only distributed bills, but also rendered them more creditworthy.[73] In case the acceptor was not itself a well-known house with a solid reputation, the discounter's guarantee provided an alternative mechanism through which borrowers could sell their bills and obtain credit in London.

To investigate this mechanism, we analyse how discounters selected their bills. The Bank of England categorised discounters into three different types: 'bankers' (all commercial banks, including mostly Anglo-foreign banks), 'brokers' (discount houses), and '[other] discounters' (a mixed bag, which included a variety of UK merchant banks and trading houses).[74] Figure 8 shows that these three types of discounters purchased similar proportions of bills drawn on small and large acceptors.[75] This means that the various discounters, whether small or large, all took part in distributing the bills accepted by the relatively less well-known, non-financial firms.

[[ INSERT **Figure 8** about here ]]

That said, the different discounter types did not all obtain their bills through the same channels. First, smaller UK trading firms and houses (included in the Bank of England's 'other discounter' category) mostly discounted bills drawn or accepted by their own trading partners. They agreed to endorse those bills because they were in a business relationship with the drawers or acceptors and knew there was a sound commercial transaction behind them.

---

[73] All the discounters in our data set served these two financial functions. Bills recorded in the Bank of England's *Discount Ledgers* had all been endorsed (and thus guaranteed) by the discounter before being resold to the Bank (the re-discounter).

[74] Of the 145 discounters in our database, 19 were 'bankers' (including three purely domestic banks, one foreign bank, and 15 Anglo-foreign banks), 19 were 'brokers' (i.e. discount houses), and 107 were '[other] discounters'.

[75] There were so many acceptors that it is not possible to identify the activities in which each was involved; also, there was no geographical variation across acceptors because all of them were based in London. Hence we can classify acceptors only in terms of their size, defined as the number of discounters who had purchased their bills. We do know that most small acceptors were merchant or industrial firms, whereas most large acceptors were established financial institutions such as commercial banks and acceptance houses.

The largest discounters were (on the one hand) discount houses and (on the other hand) Anglo-foreign banks, two types that differed in how they obtained their bills. Discount houses specialised in bill trading. They were in close contact with various acceptors and remitters of bills – London acceptance houses, banks located in foreign countries, Anglo-foreign banks, and various UK importers and exporters that accepted bills drawn on them by their trading partners – and purchased bills through these agents on a daily basis.[76]

In contrast, foreign and Anglo-foreign banks did not focus exclusively on bill discounting, and their business was geographically specialised. They maintained a large network of correspondents or branches in those areas of the world where their activities were concentrated (Jones, 1993). The correspondents shipped these banks a constant stream of bills drawn by their local customers on reputable UK financial institutions and acceptance houses as well as on smaller acceptors, especially trading and manufacturing firms. Foreign and Anglo-foreign banks discounted these bills upon their arrival in London and then either kept the bills in their respective portfolios or distributed them to other investors.[77]

[[ INSERT **Figure 9** about here ]]

The distinction between discount houses and Anglo-foreign banks is clearly apparent when examining bills re-discounted by the Bank of England. Figure 9 focuses on all discounters in our data set which had at least 10 drawers and plots the geographical concentration of their bill portfolios (as measured by the HHI)[78] against the total number of drawers with whom they were connected. The figure distinguishes between the three categories of discounters: it includes 12 'commercial banks' (including 11 Anglo-foreign banks and 1 foreign bank), 19 'discount houses', and 21 'other discounters' (merchant banks and trading houses). Our measure of geographical concentration reflects the different business models adopted by these various discounters.

---

[76] Greengrass, *Discount Market*, pp. 62-65; Vigreux, *Crédit par Acceptation*, pp. 177-78; Truptil, *British banks,* p.126; Sayers, *Gilletts*, p. 37.
[77] Greengrass, *Discount Market*, pp. 64.
[78] The index is constructed based on the geographical distribution of each discounter's drawers. We classify drawers into nine different regions depending on the city in which they were located.

Discount houses purchased bills drawn on or remitted to London agents with whom they had a relationship (either directly or through a broker). These bills could be drawn from all around the world, so the bill portfolio of a discount house was highly diversified geographically. But since Anglo-foreign banks purchased bills through their foreign correspondents, their portfolios were geographically concentrated.

[[ INSERT **Figure 10** about here ]]

Figure 10 shows the geographical composition of bills discounted by six large discounters: three discount houses (Union Discount Co, Ryder Mills & Co., National Discount Co.) and three Anglo-foreign banks (Canadian Bank of Commerce, Chartered Bank of India Australia & China, Bank of Tarapaca & Argentina). We also compare the geographical composition of these discounters' portfolios with that of the Bank of England's portfolio (i.e., the aggregate portfolio for all discounters in our data set). The discount houses' portfolios did not exhibit any specific geographical bias, and their portfolios matched the distribution of all drawers in the data set. In contrast, nearly all the bills endorsed by Anglo-foreign banks originated in the regions where those banks specialised: 89 per cent of the bills discounted by the Canadian Bank of Commerce were drawn from North America, and 85 per cent of those discounted by the Bank of Tarapaca & Argentina and by the Chartered Bank of India Australia & China originated from (respectively) Latin America and Asia/Oceania.

Through their wholesale activities, then, discounters helped reducing informational asymmetries by screening a large share of the bills on the London money market; because a discounter always endorsed the bills she distributed, their creditworthiness was enhanced by that screening. Although discounters were generally not in direct contact with drawers, they could supply information on the other intermediaries involved in a bill's origination. On the one hand, discount houses endorsed bills drawn from around the world because they knew the acceptors or remitters (either directly, or indirectly through the brokers they used and trusted). On the other hand, Anglo-foreign banks discounted bills originating from specific regions because they were remitted to them by their foreign correspondents, who had previously screened the drawers. In both cases, discounters contributed to reducing the credit risks of bills. Hence, discounters' signatures allowed for a large number of bills to be sold on the money market despite being drawn on small, unknown acceptors.

# V

This paper has presented new insights into the structure and industrial organisation of the London money market during the heyday of the first globalisation. The sterling bill market was a major pillar of the global financial system and lay at the root of the UK's financial hegemony during the years 1875-1914. Our aim is to uncover the foundations of the sterling bill's high liquidity and safety and to understand why the London money market remained so robust throughout this period.

We construct an original database that tracks the complete origination and distribution chains for all bills of exchange re-discounted by the Bank of England in 1906. Although bills re-discounted by the Bank only constituted a small portion (and were not fully representative) of all sterling bills issued, a detailed analysis of this sample provides new insights into the microstructure of the money market at the beginning of the twentieth century. We first show how borrowers from practically anywhere in the world could borrow on the London bill market. Then we describe the various mechanisms through which information asymmetries between borrowers and lenders were reduced on the money market. Market frictions and informational problems were solved thanks to the intervention of London agents (acceptors and discounters) who guaranteed the bills. All successive intermediaries involved in the origination and distribution of sterling bills contributed to produce information on the debts underlying them. This 'screening cascade' allowed unknown borrowers from even the most obscure parts of the globe to access money market investors in the world's financial capital.

Our analysis therefore reveals the crucial role of information collection – and of case-by-case screening by intermediaries – in transforming risky private debts into liquid and almost riskless money market instruments. The complex industrial organisation of the London discount market and its intermediaries' human capital and expertise were instrumental in positioning London as the world's money market and the UK as the dominant financial power during the first globalisation. The liquidity and safety of the London money market remained unquestioned until the position of bill-trading intermediaries, on which its functioning depended, was threatened by First World War's financial repercussions.[79]

---

[79] Roberts, *Saving the City*.

# APPENDIX A

# Network simulations

This appendix provides details on the network simulations presented in Section III. The purpose of these simulations is to assess whether the pattern that emerges from our data set about the number of links that drawers formed with acceptors versus discounters is not simply an artefact of the network's demography (the number of drawers, acceptors and discounters) or of the nodes' individual properties. Our data shows that drawers on the sterling bill market tended to form fewer links with acceptors than discounters. But does this pattern differ from what would have been observed if links between nodes had been formed randomly? To check this, we compare the structure of the actual network of agents with two simulated benchmarks (or null models) when link formation between nodes is randomised.

In the first benchmark, Simulation 1, we generate 100 random networks with the same number of nodes (1,361 multi-transaction drawers, 943 multi-transaction acceptors, and 119 multi-transaction discounters) and the same number of links as the actual one. We randomly recombine links between drawers on the one hand and acceptors/discounters on the other hand. This first simulation is based on a simple conditional U/L distribution[80] where the number of nodes and ties is fixed. In this scenario, each acceptor (resp. discounter) has the same likelihood as any other acceptor (resp. discounter) to form a link with a drawer. In other words, each acceptor/discounter appears on roughly the same number of bills.[81] Simulation 1 allows us to visualise the distribution of drawers' acceptor-to-discounter ratios had links between nodes been generated in a purely random fashion.

---

[80] Wasserman and Faust, *Social network analysis*.
[81] In order to produce this scenario, we divide the total number of transactions involving the 1,361 multi-transaction drawers (6,715 transactions) by the number of acceptors (943) and the number of discounters (119). We then create two columns: one listing all acceptors and one listing all discounters. In the acceptors' (respectively, discounters') column, each acceptor (respectively, discounter) appears as many times as in the observed network (rounded up to the nearest integer value). Thus, since the observed drawer-to-acceptor ratio is 7.12, each acceptor appears 8 times in the acceptors' column. Since the observed drawer-to-discounter ratio is 56.43, each discounter appears 57 times in the discounters' column. We then produce simulations through a process of column building by randomly associating drawers in the original drawers' column with acceptors (respectively, discounters) in the newly-constructed acceptors' (respectively, discounters') column. Each recombination of the 6,715 rows constitutes one simulated network. We repeat this procedure 100 times in order to produce 100 different simulated networks.

In our second benchmark, Simulation 2, we generate 100 random networks with the same number of nodes, the same number of links, and the same degree distribution (the total number of links of each node) as the actual one. Simulation 2 (known as 'degree preserving randomisation' in social network analysis) therefore better accounts for acceptors/discounters differing in their respective abilities to form links with drawers. Thus, in the language of social network analysis, we account for various nodes having different 'relational capacities'. In the simulated networks, each node also has the same total number of links (or degree) as in the actual data. Formally, this random model is based on a simple U/L distribution with specified in-degree and out-degree.[82] This means that each acceptor/discounter has the same likelihood of forming a link with a drawer in the simulated network as in the observed network.[83] Simulation 2 allows us to check whether the pattern we observe in the data about drawers' acceptor-to-discounter ratio does not simply arise from the fact that various acceptors/discounters had different relational capacities (total number of links).

Figure A.1 plots the frequency distribution of drawers' acceptor-to-discounter ratio in the actual data (white bars) together with the frequency distribution of the same variable in random networks generated according to our two simulations.[84] For each of the 100 networks generated through

---

[82] Wasserman and Faust, *Social network analysis*.

[83] This scenario is also produced through a process of column building. Each actor's likelihood to appear is unchanged with respect to the observed data. In order to produce a simulated network, we now simply recombine the original acceptors' and discounters' columns while keeping the original drawers' column fixed. Hence, each acceptor and discounter has as many transactions in the simulated network as in the observed one, while the pattern of these transactions is redefined randomly. We then repeat this procedure 100 times in order to produce 100 simulated networks.

[84] In order to check if the actual and simulated distributions of drawers' acceptor-to-discounter ratios are statistically-significantly different from each other, we perform a two-sample Kolmogorov-Smirnov (KS) test. The KS test is a non-parametric test which allows comparing two cumulative distribution functions under the null hypothesis that the two distributions are equal. The test compares the unaggregated frequency distribution of our actual network with the unaggregated frequency distribution of each random network we generate. For each simulated network, the test shows p-values close to 0, indicating that the frequency distributions of the actual and random networks are statistically-significantly different from each other. The KS statistic represents the maximum distance between two compared distributions. This statistic ranges from 0.4909 to 0.5018 when comparing the actual distribution of the acceptor-to-discounter ratio to the 100 distributions generated through Simulation 1. It ranges from 0.4569 to 0.4772 when comparing the actual distribution to the 100 distributions generated through Simulation 2. This indicates that all simulated distributions are far apart from the actual distribution (even though, as expected, the distance between actual and simulated distributions is smaller in the case of Simulation 2).

simulation 1 (resp., simulation 2), a grey tilde (resp., a grey line) indicates the number of drawers with a given acceptor-to-discounter ratio.

[[ INSERT **Figure A.1** about here ]]

In the actual network, most drawers have an acceptor-to-discounter ratio that is strictly less than 1 (i.e., they have fewer acceptors than discounters). But if links between nodes had been generated randomly, then (a) an overwhelming majority of drawers would have displayed an acceptor-to-discounter ratio of exactly 1 – that is, an equal number of acceptors and discounters – and (b) a higher proportion of drawers would have exhibited a ratio strictly greater than 1 (more acceptors than discounters) than strictly less than 1 (more discounters than acceptors). This outcome reflects that the network includes more acceptors than discounters. It is also worth noting that more outliers appear in the observed network than in the simulated ones, which suggests that the determinants of link formation behaviour varied greatly for different actors.

Finally, we investigate the extent to which acceptors (discounters) had drawers in common. Figure A.2 reports the frequency distribution of acceptors (panel A) and discounters (panel B) according to the percentage (x) of fellow acceptors/discounters with whom they shared at least one drawer.[85] In each case, we report the observed distribution (white bars) in the actual network as well as the distributions in the simulated networks obtained through simulation 1 (grey tildes) and simulation 2 (grey lines).

[[ INSERT **Figure A.2** about here ]]

In the observed distributions we can see that acceptors were less likely than discounters to share drawers with their peers: 40 per cent of the acceptors in our data set did not share any of their drawers with other acceptors, although more than 75 per cent of the discounters shared at least one drawer with other discounters. No acceptor shared a drawer with more than 13 per cent of the other acceptors, but a sizable group of discounters shared drawers with more than 40 per cent of their fellow discounters.

---

[85] We equally perform a KS test to compare the actual and simulated distributions of these variables. For the acceptors' 'shared drawers' variable, we find p-values close to zero indicating that the actual and simulated distributions are statistically significantly different from each other; the maximum distance between the actual distribution and the 100 distributions generated ranges from 0.7400 to 0.7665 for Simulation 1, and from 0.2605 to 0.3148 for Simulation 2. In the case of the discounters' 'shared drawers' variable, p-values are again close to zero; the maximum distance between the actual distribution and the 100 distributions generated is always equal to 1.000 for Simulation 1, while it ranges from 0.2206 to 0.3379 for Simulation 2.

If links between drawers and acceptors had been formed randomly, then acceptors would (on average) have shared more drawers among them than they actually did (Figure A.2, panel A). This means that the tendency of acceptors not to share drawers, which we observe in the data, is due not to our network's structural characteristics but rather to structural factors in the formation of links between drawers and acceptors. The low amount of sharing observed among acceptors in the actual data strongly suggests that they held private information on drawers.

In panel B of the figure we see that discounters were divided into two groups. The small discounters in our data set shared, on average, fewer drawers than predicted by the simulations; however, large discounters shared as many drawers as predicted for the case of randomly formed links between drawers and discounters. These results indicate that, unlike the acceptors, large discounters did *not* hold proprietary information on the drawers.

**Figure A.1.** Acceptor/discounter ratios: Observed versus simulated

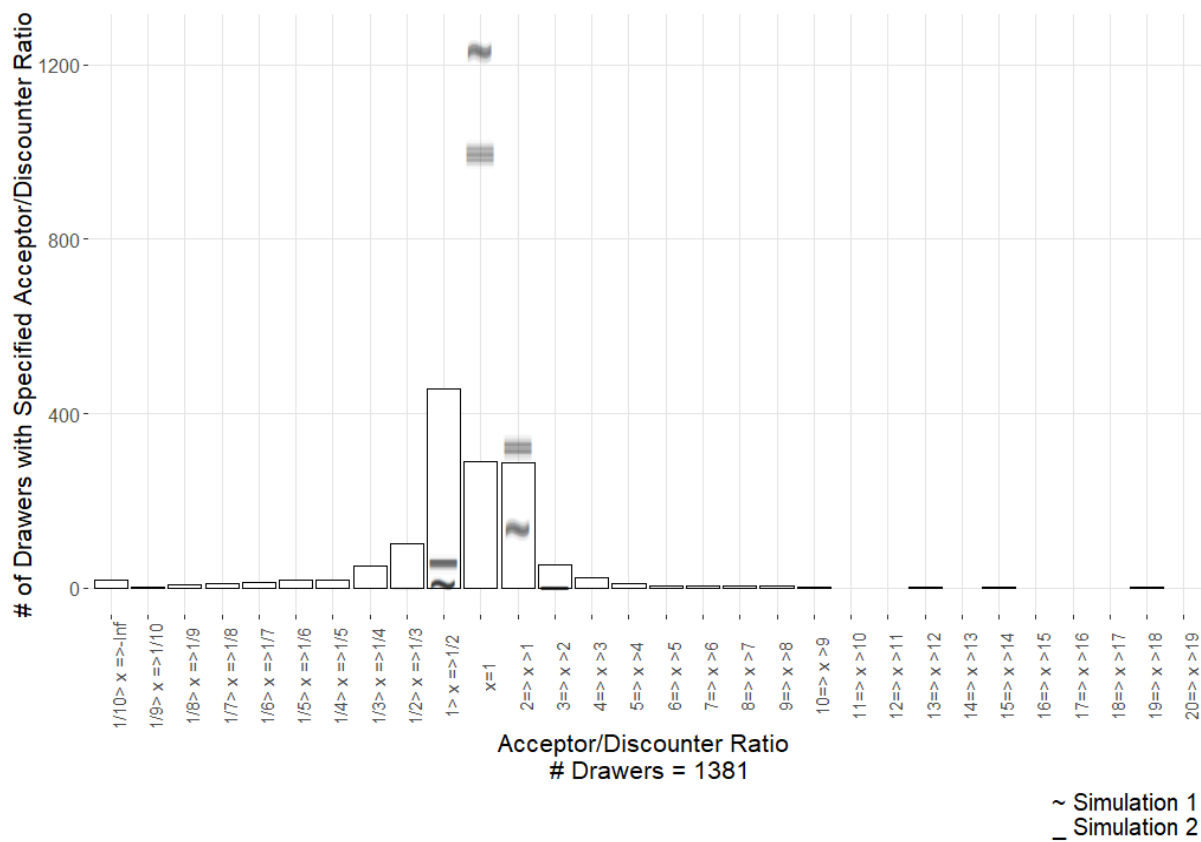

*Notes:* The figure plots the frequency distribution of the 1,381 multi-transaction drawers according to their acceptor-to-discounter ratio (denoted by x on the horizontal axis) in the observed network (white bars) as well as in simulation 1 (grey tildes) and simulation 2 (grey lines). Drawers for whom x < 1 are linked to more discounters than acceptors. Drawers for whom x = 1 are linked to as many discounters as acceptors. Drawers for whom x > 1 are linked to more acceptors than discounters. Each tilde (line) indicates the number of drawers in the corresponding decile for each of the 100 simulations (we do not report a tilde or line when there is no drawer in the corresponding decile).

**Figure A.2.** Shared drawers: Observed versus simulated
**Panel A:** Acceptors

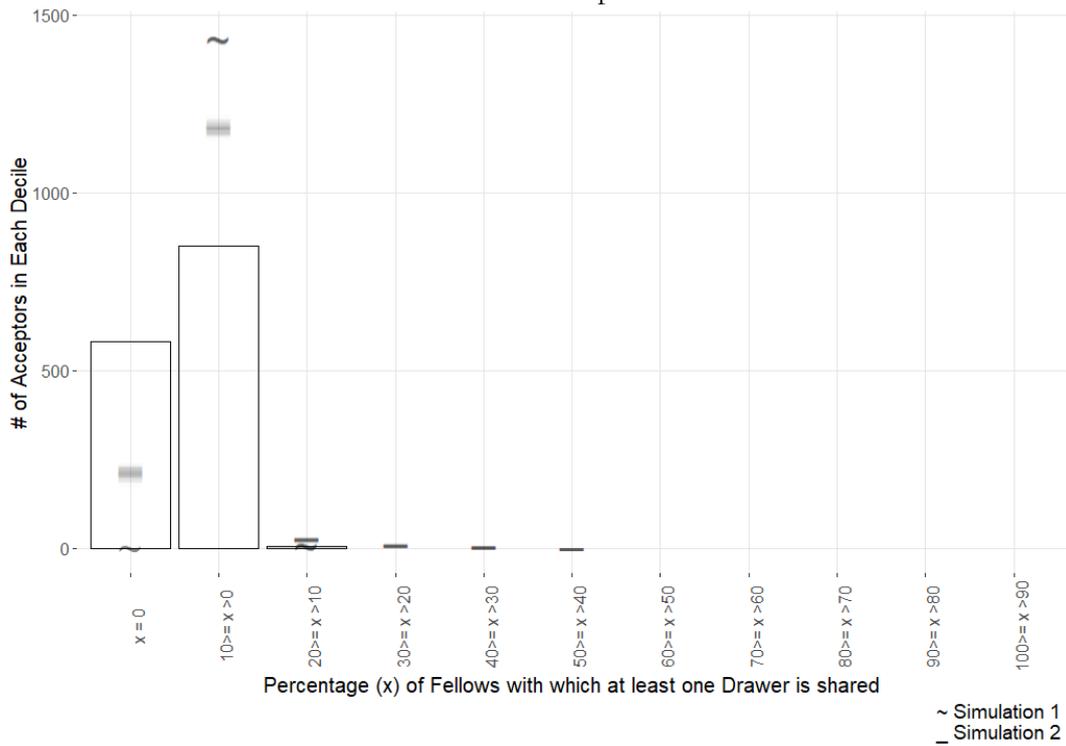

**Panel B:** Discounters

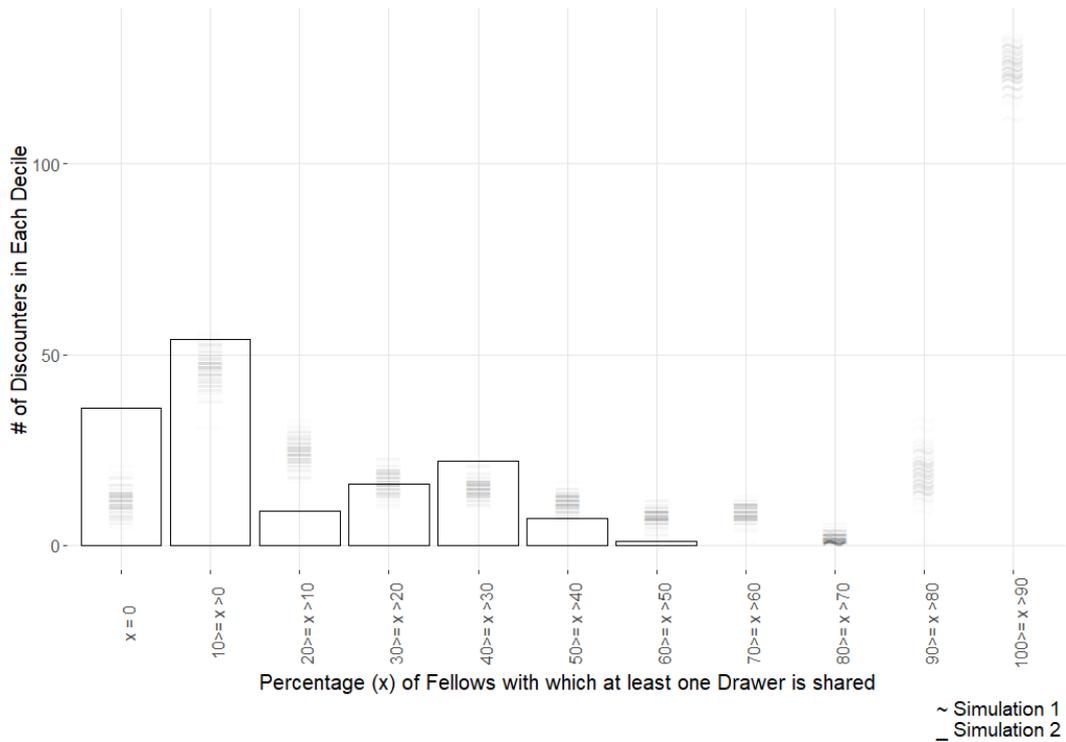

*Notes:* This figure plots the frequency distribution of acceptors (panel A) and discounters (panel B) according to the percentage (x) of fellow acceptors/discounters with whom they share at least one drawer in the observed network (white bars) as well as in simulation 1 (grey tildes) and simulation 2 (grey lines). Acceptors/discounters for which x = 0 do not share any drawer with any of their peers, whereas acceptors/discounters for which x = 100 share at least one drawer (not necessarily the same) with all other fellow acceptors/discounters. Each tilde (line) indicates the number of drawers in the corresponding decile for each of the 100 simulations (we do not report a tilde or line when there is no drawer in the corresponding decile).

# APPENDIX B

## Network analysis – Descriptive statistics

This appendix provides descriptive statistics on the network of agents involved in the origination and distribution of bills re-discounted by the Bank of England in 1906. We also provide similar statistics for simulated, random networks (null models, see text and appendix A).

[[ INSERT **Figure B.1** about here ]]

As shown in figure B.1, our relational unit is a *triad* involving three roles (drawer-acceptor-discounter). A bill always involves a direct link between its drawer and its acceptor as well as a direct link between its acceptor and its discounter. This means that an indirect relationship also exists between its drawer and its discounter (through the acceptor). Formally, the relation between agents forming a bill's triad is a *compound relation*. This specific feature of our network has implications for the interpretation of standard descriptive indicators used in network analysis. These indicators assume that the relational unit of analysis is a *dyad* (rather than a triad). In the case of our network, this means that standard measures do not account for the indirect relationship that exists between drawers and discounters. This makes the interpretation of standard indicators less straightforward. For example, since an acceptor always lays in the middle of each triad, descriptive measures based on the degree or paths (*geodesic distances*) will tend to overestimate the structural importance of acceptors in the network. The descriptive network measures presented below are therefore not directly comparable to those produced for modern banking networks.

[[ INSERT **Table B.1** about here ]]

Table B.1. reports summary statistics about the importance (or centrality) of the various nodes in the network. A first measure of centrality is the *degree* which corresponds to the total number of links each node has. A node's *in-degree* is equal the total number of links received. In our network, the in-degree corresponds to the total number of dyadic relationships in which a given node plays the role of acceptor of bills drawn by another node or of discounter of bills

accepted by another node. A node's *out-degree* is equal to the number of links sent. In our network, a node's out-degree is equal to the total number of dyadic relationships in which a given node plays the role of drawer of bills accepted by another node or of guarantor of bills discounted by another node.

A second way of measuring a node's centrality is through its *distance* or *path* to other nodes (the number of steps the node needs to reach other nodes in the network). There exist two standard distance-based measures of node centrality. First, *closeness* measures how close a node is from all other nodes in the network. It is defined as the inverse of the sum of the lengths of the *shortest paths* (or *geodesic distances*) between a given node and all other nodes in the network. Closeness is usually associated to the node's capacity to impact the rest of the network. A second distance-based indicator is *betweenness* which measures the extent to which a given node acts as an intermediary between other nodes in the network. It is computed by calculating the number of times a node lays on the shortest path between two other nodes in the network (as a fraction of the total number of shortest paths between all pairs of nodes in the network). Betweenness therefore measures a given node's ability to control relationships between other nodes.

Another standard indicator is *eigenvector centrality*, which measures a node's centrality through the centrality of the other nodes with which it is linked. Eigenvector centrality is a weighted degree centrality measure as it weighs a node's links according to their importance.

Table B.1 reports summary statistics for each of these indicators in the actual network and in the random network$q^2$s generated through Simulations 1 and 2.

[[ INSERT **Table B.2** about here ]]

In table B.2., we also report additional descriptive statistics on the structural properties of the actual and random networks. *Clustering* (sometimes called *transitivity*) is an indicator of the presence of groups of nodes exhibiting relatively more dense interconnections between them. It is computed as the share of all possible sets of three nodes that actually display a complete interconnection. *Closeness* (respectively, *betweenness*) *centralization* is an indicator of the degree of inequality in closeness (respectively, betweenness) among nodes in the entire network. Centralization is computed as the sum of the differences between the most central node's centrality indicator and that of each other node in the network. The values range from 0 to 1, where values close to 0 indicate low inequality across nodes. A *component* is a set of nodes which are all connected to each other by a path. The *main component* refers to the share of nodes which belong to the largest component in the network. A main component of 100 therefore indicates that all nodes in the network are directly or indirectly connected with each other. Finally, the network's *average path distance* is defined as the average geodesic distance (or number of steps) between all pairs of nodes. This measure provides information about how relatively close nodes are.

Table B.2 reports the value of each of these indicators for the actual network as well as for the random networks. In the case of random networks, the table reports the range and mean of each measure across the 100 generated networks. Overall, the actual network exhibits higher clustering than the random ones, suggesting that interconnections do not emerge by chance. The actual network also exhibits lower closeness centralization, but higher betweenness centralization than randomly generated networks. This suggests that there were relatively few highly central agents (in terms of closeness) on the bill market, but a few agents nevertheless exhibited strong ability to control relationships between other nodes. The actual network's main component takes a value of 99.3, indicating that only 0.7 per cent of nodes were isolated: this is only slightly lower than the values found in simulated networks. Finally, the actual network's average path length is 4.82, which is close to the average path length in random networks generated through degree preserving randomization (Simulation 2).

**Figure B.1.** Links between a triad of nodes appearing on a bill of exchange

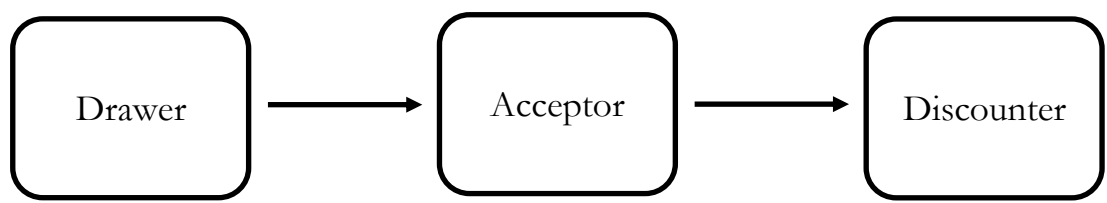

*Notes:* This figure provides a schematic representation of the links between a triad of nodes (or agents) appearing on a same bill of exchange.

**Table B.1.** Node centrality - Summary statistics

|  | In-degree | | | Out-degree | | | All-degree | | |
|---|---|---|---|---|---|---|---|---|---|
|  | Min | Mean | Max | Min | Mean | Max | Min | Mean | Max |
| Actual | 0 | 2.10 | 357 | 0 | 2.10 | 50 | 1 | 4.20 | 375 |
| Simulation 1 | 0 | 2.10 | 48 | 0 | 2.10 | 10 | 1 | 4.20 | 54 |
| Simulation 2 | 0 | 2.10 | 357 | 0 | 2.10 | 50 | 1 | 4.20 | 375 |
|  | Closeness | | | Betweenness | | | Eigenvector centrality | | |
|  | Min | Mean | Max | Min | Mean | Max | Min | Mean | Max |
| Actual | 0.00 | 0.03 | 0.03 | 0.00 | 0.00 | 0.02 | 0.00 | 0.00 | 0.19 |
| Simulation 1 | 0.17 | 0.21 | 0.30 | 0.00 | 0.00 | 0.01 | 0.00 | 0.01 | 0.14 |
| Simulation 2 | 0.00 | 0.14 | 0.43 | 0.00 | 0.00 | 0.01 | 0.00 | 0.00 | 0.34 |

*Notes:* This table reports summary statistics of various node centrality indicators for the actual network of money market agents as well as for the random networks generated through Simulations 1 and 2 (see text and appendix A for details on the simulations). For each simulation, Min (Max) correspond to the minimum (maximum) value of the indicator across all nodes in the 100 simulated random networks and Mean corresponds to the mean value of each indicator across all 100 networks generated. By construction, each node in the networks generated through Simulation 2 has the same degree as in the actual network. *Sources*: see text.

**Table B.2.** Network's structural properties – Descriptive statistics

|  | Actual | Simulation 1 | | | Simulation 2 | | |
|---|---|---|---|---|---|---|---|
|  |  | Min | Mean | Max | Min | Mean | Max |
| Clustering | 0.038 | 0.002 | 0.002 | 0.003 | 0.018 | 0.020 | 0.021 |
| Closeness centralization | 0.002 | 0.152 | 0.161 | 0.178 | 0.009 | 0.089 | 0.317 |
| Betweenness centralization | 0.017 | 0.004 | 0.006 | 0.009 | 0.005 | 0.009 | 0.013 |
| Main component | 99.3 | 100.0 | 100.0 | 100.0 | 99.7 | 99.9 | 100.0 |
| Average path length | 4.819 | 6.194 | 6.973 | 7.726 | 4.047 | 4.253 | 4.496 |

*Notes:* This table reports descriptive statistics on the structural properties of the actual network of money market agents as well as of the random networks generated through Simulations 1 and 2 (see text and appendix A for details on the simulations). For each simulation, the table reports the minimum, maximum, and mean value of each indicator across all 100 simulated networks. *Sources*: see text.

# References

Accominotti, O., 'London Merchant Banks, the Central European Panic, and the Sterling Crisis of 1931', *Journal of Economic History*, 72 (2012), pp. 1-43.

Accominotti, O., 'International banking and transmission of the 1931 financial crisis', *The Economic History Review*, 72 (2019), pp. 260-285.

Accominotti, O., and Ugolini S., 'International Trade Finance from the Origins to the Present: Market Structures, Regulation and Governance', in Brousseau, E., Glachant, J.-M., and Sgard, J. (eds.), *Oxford Handbook on International Economic Governance*, (Oxford, 2019).

Atkin, J., *The Foreign Exchange Market of London* (New York, 2005).

Baster, A. S. J., *The International Banks* (London, 1935).

Berger, Allen N. and Udell G. F. (1995), 'Relationship banking and lines of credit in small firm finance', *The Journal of Business*, 68 (1995), pp. 351-381.

Bignon, V., Flandreau M., and Ugolini S., 'Bagehot for Beginners: The Making of Lending-of-Last-Resort Operations in the Mid-Nineteenth Century', *Economic History Review*, 65 (2012), pp. 580-608.

Boot, A. W. A., 'Relationship Banking: What Do We Know?', *Journal of Financial Intermediation*, 9 (2000), pp. 7-25.

Boot, A. W. A. and Thakor, A. V. (2000), 'Can relationship banking survive competition?', *The Journal of Finance*, 55 (2000), pp. 679-713.

Cassis, Y. (2006), *Capitals of capital : a history of international financial centres, 1780-2005*, (Cambridge, 2006).

Chapman, S., *The Rise of Merchant Banking* (London, 1984).

Chinazzi, M., Fagiolo G., Reyes J. A., and Schiavo S. (2013), 'Post-mortem examination of the international financial network', *Journal of Economic Dynamics and Control*, 37 (2013), pp. 1692–1713.

Clare, G., *The A B C of the foreign exchanges: a practical guide* (London, 1911).

Cleaver, G. and Cleaver, P., *The Union Discount: a centenary album* (London, 1985).

Committee on Finance and Industry, *Minutes of Evidence* (London, 1931).

Craig, B., and von Peter G., 'Interbank tiering and money center banks', *Journal of Financial Intermediation*, 23 (2014), pp. 322-347.

de Roover, R. (1953), *L'évolution de la lettre de change, XIVe-XVIIIe siècles* (Paris, 1953).

Diamond, D. W., 'Financial intermediation and delegated monitoring', *The Review of Economic Studies*, 51 (1984), pp. 393-414.

Diamond, D. W., 'Monitoring and reputation: the choice between bank loans and directly placed debt', *Journal of Political Economy*, 99 (1991), pp. 689-721.

Eichengreen, B. and Flandreau, M., 'A Century and a Half of Central Banks, International Reserves, and International Currencies', in Bordo, M. D., Eitrheim, O., Flandreau, M., and Qvigstad, J.F., (eds.) *Central Banks at a Crossroads: What Can We Learn from History?*, (New York, 2016), pp. 280-318.

Flandreau, M. and Jobst, C., 'The Ties That Divide: A Network Analysis of the International Monetary System, 1890–1910', *The Journal of Economic History*, 65 (2005), pp. 977-1007.

Flandreau, M. and Mesevage, G., 'The separation of information and lending and the rise of rating agencies in the USA (1841-1907)', *Scandinavian Economic History Review*, 62 (2014), pp. 213-242.

Flandreau, M. and Ugolini S. (2013), 'Where It All Began: Lending of Last Resort and Bank of England Monitoring during the Overend-Gurney Panic of 1866', in Bordo, M. D. and Roberds W. (eds.), *The Origins, History, and Future of the Federal Reserve: A Return to Jekyll Island*, (New York, 2013), pp. 113-161.

Fletcher, G. A., *The discount houses in London: principles, operations and change* (London, 1976).

Ghatak, M. and Guinanne T. W., 'The Economics of Lending with Joint Liability: Theory and Practice', *Journal of Development Economics*, 60 (1999), pp. 195-228.

Gillett Brothers Discount Company Ltd, *The Bill on London: Or, the Finance of Trade by Bills of Exchange* (London, 1952).

Goschen, G. J., *The theory of the foreign exchanges* (London, 1876).

Greengrass, H. W., *The Discount Market in London* (London, 1931).

Herger, N., 'Interest-parity conditions during the era of the classical gold standard (1880-1914) - evidence from the investment demand for bills of exchange in Europe', *Swiss Journal of Economics and Statistics*, 154 (2018), pp. 1-12.

Hawtrey, R. G., *Currency and Credit* (London, 1930).

Iori, G., De Masi, G., Precup, O.V., Gabbi, G., and Caldarelli G., 'A network analysis of the Italian overnight money market', *Journal of Economic Dynamics and Control*, 32 (2008), pp. 259–278.

Jacobs, L. M., *Bank Acceptances*, United States Senate Sixty-First Congress, Second Session (document no 569) (Washington DC, 1910).

Jansson, W., *The Finance-Growth Nexus in Britain, 1850-1913*, unpublished PhD dissertation, (University of Cambridge, 2018).

Jobst, C. and Ugolini S., 'The Coevolution of Money Markets and Monetary Policy, 1815-2008', in Bordo, M. D., Eitrheim, O., Flandreau, M., and Qvigstad, J. F. (eds.), *Central Banks at a Crossroads: What Can We Learn from History?* (New York, 2016), pp. 145-194.

Jones, G. G., *British Multinational Banking, 1830-1990* (Oxford, 1993).

King, W. T. C., *History of the London Discount Market* (London, 1936).

Keynes, J. M., *A Treatise on Money: The Applied Theory of Money*, in Johnson, E. and Moggridge, D. (eds.), *The Collected Writings of John Maynard Keynes*, Cambridge, 6 (1978 [1930]).

Kynaston, D., *The City of London, volume 2. Golden Years, 1890-1914* (London, 1994).

Martinez-Jaramillo, S., Alexandrova-Kabadjova, B., Bravo-Benitez, B. and Solórzano-Margain, J. P., 'An empirical study of the Mexican banking system's network and its implications for systemic risk', *Journal of Economic Dynamics and Control*, 40 (2014), pp. 242-265.

Mollan, S. and Michie R. C., 'The City of London as an International Commercial and Financial Center since 1900', *Enterprise and Society*, 13 (2012), pp. 538-596.

Michie, R. C., *British Banking: Continuity and Change from 1694 to the Present* (Oxford, 2016).

Nier, E., Yang, J., Yorulmazer, T. and Alentorn, A., (2007), 'Network models and financial stability', *Journal of Economic Dynamics and Control*, 31(6) (2007), pp. 2033-60.

Nishimura, S., *The Decline of Inland Bills of Exchange in the London Money Market, 1855-1913* (Cambridge, 1971).

Petersen, M. A. and Rajan, R. G. (1994), 'The benefits of lending relationships: evidence from small business data', *The Journal of Finance*, 49 (1994), pp. 3-37.

Rajan, R. G., 'Insiders and Outsiders: The Choice Between Informed and Arm's-Length Debt', *The Journal of Finance*, 47 (1992), pp. 1367-1400.

Roberts, R., *Schröders: Merchants and Bankers* (London, 1992).

Roberts, R., *Saving the City: The Great Financial Crisis of 1914* (Oxford, 2013).

Santarosa, V. A., 'Financing Long-Distance Trade: The Joint Liability Rule and Bills of Exchange in Eighteenth-Century France' *Journal of Economic History*, 75 (2015), pp. 690-719.

Sayers, R. S., *Gilletts in the London Money Market, 1867-1967* (Oxford, 1968).

Scammell, W. M., *The London Discount Market* (London, 1968).

Sharpe, S. A., 'Asymmetric Information, Bank Lending and Implicit Contracts: A Stylized Model of Customer Relationships', *The Journal of Finance*, 45 (1990), pp. 1069-1087.

Spalding, W. F. (1915), *Foreign exchange and foreign bills in theory and in practice* (London, 1915).

Stein, J. C. (2002), 'Information Production and Capital Allocation: Decentralized versus Hierarchical Firms', *The Journal of Finance*, 57 (2002), pp. 1891-1922.

Stiglitz, J. E. and Weiss A., 'Credit rationing in markets with imperfect information', *The American Economic Review*, 71 (1981), pp. 393-410.

Truptil, R. (1936), *British banks and the London money market* (London, 1936).

Ugolini, S., 'Liquidity Management and Central Bank Strength: Bank of England Operations Reloaded, 1889-1910', Norges Bank Working Paper, 10/2016.

Vigreux, P.-B., *Le Crédit par Acceptation. Paris Centre Financier* (Paris, 1932).

Warburg, P. M., *The Discount System in Europe*, United States Senate Sixty-First Congress, Second Session (document no 402) (Washington DC, 1910).

Wake, J., *Kleinwort Benson: the history of two families in banking* (Oxford, 1997).

Wasserman, S. and Faust K., *Social network analysis: Methods and applications* (Cambridge, 1994).

Withers, H., *The meaning of money* (New York, 1920.

**Table 1.** Ranking of London acceptance houses in 1906

|  | Our database | | Jansson (2018, p. 269) | |
| --- | --- | --- | --- | --- |
|  | Rank | Number of discounters | Rank | Amount accepted (million pounds) |
| Kleinworts | 1 | 50 | 1 | 11.9 |
| Schröders | 2 | 45 | 2 | 10.3 |
| Barings | 3 | 43 | 4 | 6.7 |
| Brandts | 4 | 42 | 3 | 6.9 |
| Brown Shipley | 5 | 39 | 5 | 4.5 |
| Rothschilds | 6 | 33 | 6 | 3.1 |
| Hambros | 7 | 32 | 8 | 2.1 |
| Morgan Grenfell | 8 | 26 | 7 | 2.2 |
| Gibbs | 9 | 19 | 9 | 0.9 |

*Notes:* This table compares the ranking – in terms of their number of discounters in 1906 – of nine of our data set's acceptance houses with the ranking of the same houses established by Jansson, *Finance-Growth Nexus in Britain*, p. 269, based on their amount of outstanding accepted bills at the end of 1906, retrieved from the respective houses' archival records.

*Sources:* See text.

**Table 2. Comparison of acceptors in Gilletts' total discounts and re-discounts to the Bank of England, 1906**

| Type of acceptor | Share in Gilletts' total discounts | Share in Gilletts' re-discounts to BoE |
|---|---|---|
| Merchant banks | 49.5% | 32.0% |
| British clearing banks | 9.8% | 16.4% |
| Anglo-foreign banks | 11.9% | 30.9% |
| Foreign banks | 13.1% | 0.0% |
| Other (non-financial firms) | 15.7% | 20.7% |
| All | 100.0% | 100.0% |

*Notes:* This table compares the share of various types of acceptors in Gilletts' 1906 total portfolio of discounted bills and in the portion of that portfolio which the firm re-discounted to the Bank of England.
*Sources:* See text.

**Table 3. Correlation between the size of acceptors in Gilletts' and Bank of England's bill portfolios, 1906**

| Type of acceptor | Coefficient of correlation |
| --- | --- |
| All | 0.52*** |
| All (excluding foreign banks) | 0.62*** |
| Merchant banks | 0.71*** |
| British clearing banks | 0.62*** |
| Anglo-foreign banks | 0.55*** |
| Foreign banks | 0.75*** |
| Other | 0.64*** |

*Notes:* This table shows the coefficient of correlation between the size of acceptors in Gilletts' 1906 bill portfolio and their size in the Bank of England's aggregate portfolio of re-discounted bills. The size of a given acceptor in Gilletts' bill portfolio is measured through the total amount of bills carrying this acceptor's signature. The size of a given acceptor in the Bank of England's portfolio is measured through its total number of discounters. The correlation is shown for all acceptors in the sample as well as for different categories of acceptors. ***: significant at the 1 per cent level.

*Sources:* See text.

**Table 4.** Acceptors and discounters per drawer

| Panel A: Number of Acceptors/Discounters per Drawer | | | | |
|---|---|---|---|---|
| | Mean | SE | Max | Min |
| *Acceptors* | | | | |
| No. of acceptors per drawer | 2.83 | (0.08) | 38 | 1 |
| % of all acceptors | 0.20 | (0.01) | 2.64 | 0.07 |
| *Discounters* | | | | |
| No. of discounters per drawer | 3.33 | (0.08) | 36 | 1 |
| % of all discounters | 2.29 | (0.05) | 24.83 | 0.69 |

| Panel B: Repartition of Drawers | | | | |
|---|---|---|---|---|
| No. of transactions involved | Total drawers | Discounters > Acceptors | Discounters = Acceptors | Discounters < Acceptors |
| *All > 1* | | | | |
| Observed | 1,381 | 50.25% | 21.07% | 28.67% |
| Simulation 1 | 1,381 | 0.79% | 89.28% | 9.93% |
| Simulation 2 | 1,381 | 4.26% | 72.21% | 23.53% |
| *2* | | | | |
| Observed | 558 | 47.67% | 26.70% | 25.62% |
| Simulation 1 | 558 | 0.10% | 99.06% | 0.85% |
| Simulation 2 | 558 | 1.16% | 94.65% | 4.18% |
| *3* | | | | |
| Observed | 239 | 44.35% | 30.54% | 25.10% |
| Simulation 1 | 239 | 0.34% | 97.20% | 2.46% |
| Simulation 2 | 239 | 3.42% | 85.26% | 11.32% |
| *4* | | | | |
| Observed | 158 | 58.86% | 13.92% | 27.21% |
| Simulation 1 | 158 | 0.55% | 94.44% | 5.01% |
| Simulation 2 | 158 | 5.70% | 73.00% | 21.30% |
| *5–9* | | | | |
| Observed | 286 | 55.59% | 12.23% | 32.16% |
| Simulation 1 | 286 | 1.47% | 84.39% | 14.14% |
| Simulation 2 | 286 | 9.50% | 46.71% | 43.79% |
| *10+* | | | | |
| Observed | 140 | 50.00% | 8.57% | 41.42% |
| Simulation 1 | 140 | 3.24% | 40.91% | 55.85% |
| Simulation 2 | 140 | 5.71% | 11.67% | 82.62% |

*Notes:* This table focuses on multi-transaction drawers, or those whose names appear on at least two non-identical bills in our data set. Panel A reports the mean, the standard error, and the maximum and minimum number of multi-transaction drawers per acceptor and discounter. Panel B (Observed) displays the share of multi-transaction drawers with more discounters than acceptors, with as many discounters as acceptors, and with fewer discounters than acceptors. The repartition is shown for all drawers in our data set who have more than one different bill (All > 1) as well as for drawers with different numbers of bills. Panel B (Simulation 1 and Simulation 2) also reports the same repartition in two simulated networks generated on the assumption that links between drawers and acceptors/discounters were formed randomly (for details, see text and the Appendix).

*Source:* Bank of England's *Discount Ledgers*.

**Table 5. Market concentration in accepting and discounting**

|  | Acceptors | Discounters |
|---|---|---|
| *HHI* | 116.05 | 422.53 |
| *Highest market penetration* | 9.14% | 19.84% |
| *Market share of top actors* | | |
| Top 3 | 11.21% | 23.07% |
| Top 5 | 16.63% | 33.72% |
| Top 10 | 28.10% | 55.87% |
| Top 15 | 35.95% | 71.69% |

*Notes:* This table presents several indicators of market concentration for acceptors and discounters of bills re-discounted by the Bank of England: the HH index, the highest market penetration, and the market share of the top 3, top 5, top 10, and top 15 acceptors and discounters. See text for details on these indicators.

*Source:* Bank of England's *Discount Ledgers* (see text).

**Figure 1.** Geographical location of borrowers (drawers) on the London bill market

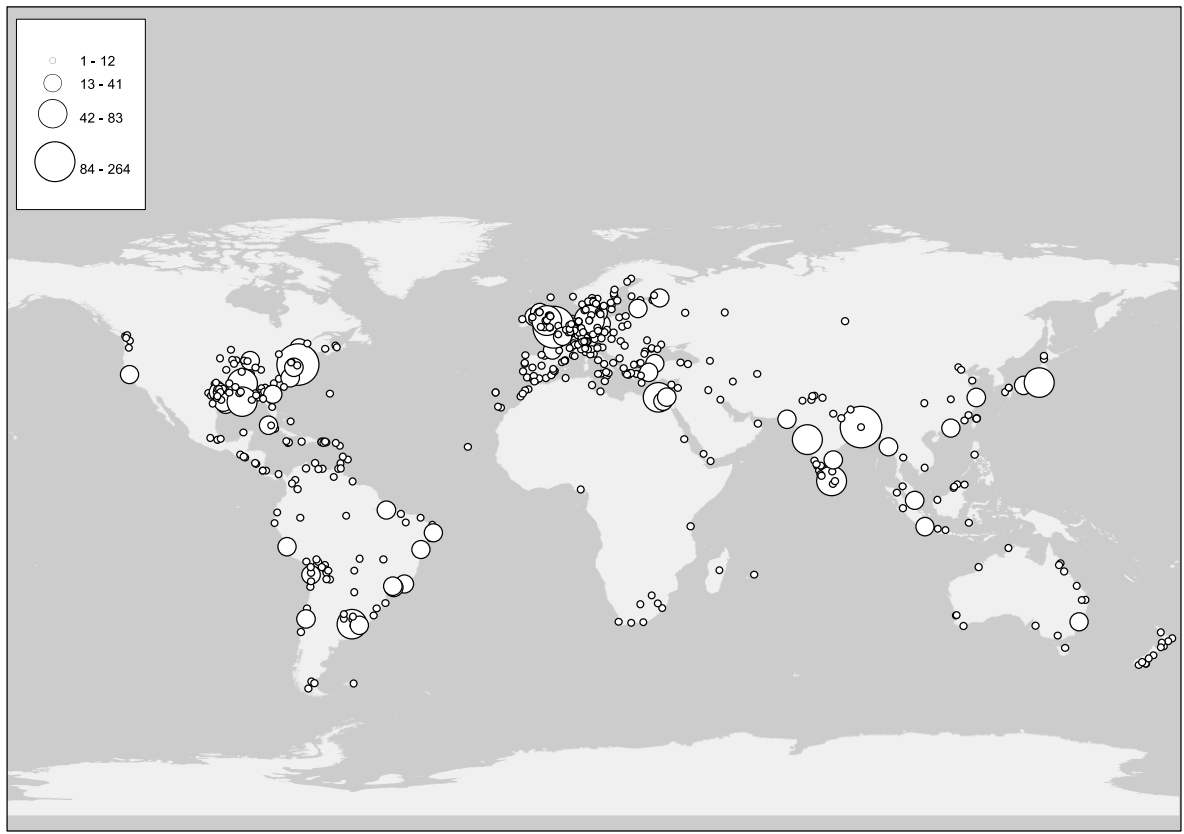

*Note:* This map shows the geographical location (at the city level) of all drawers of sterling bills in our data set.
*Source:* Bank of England's *Discount Ledgers* (see text.)

**Figure 2.** Example of a commercial transaction financed by a sterling bill of exchange

**Panel A:** Operations at issue

**Panel B:** Operations at maturity

*Source:* Authors' schematic representation of transactions described by contemporaries (e.g., Gillett Brothers, *Finance of Trade by Bills of Exchange*)

**Figure 3.** Bill of exchange, 1910

|  | Moscow, 10 August 1910 |
| --- | --- |
| ACCEPTED 15 AUGUST 1910<br>Payable at the<br>London County & Westminster Bank<br>*Kleinwort Sons & Co.* | For £3,000 |

    Three months after date pay against this Bill of Exchange to our order the Sum of <u>Three Thousand Pounds Sterling</u> Value in account and place it to account and charge it to account as advised by

Messrs Kleinwort Sons & Co.                   *Société L. Bauer & Co.*
    <u>London E.C.</u>                          *Le Directeur-Gérant: C. Bauer*
ac Fenchurch Street

---

Pay to the order of:
Banque de Commerce de l'Azow-Don
10 August 1910
*Société L. Bauer & Co.*

For us at the order of:
The Union Discount Company of London Ltd., London
10 August 1910
*Azow-Don Commerzbank*

*Notes:* The figure's upper (resp. lower) portion transcribes the bill's front (resp. back) side. Text in italics corresponds to signatures.
*Source:* London Metropolitan Archives, CLC/B/140/KS04/13/02/006.

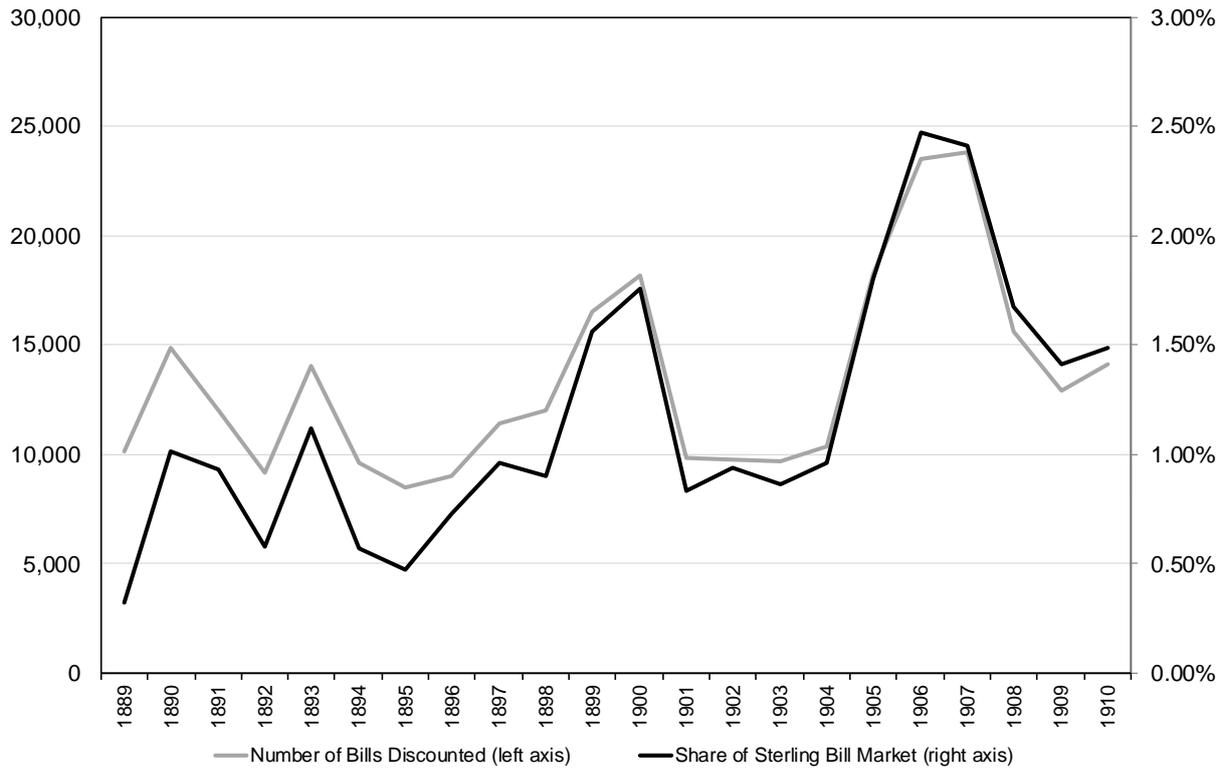

**Figure 4.** The Bank of England's re-discount policy, 1889-1910

*Note:* This figure displays the total number of bills re-discounted by the Bank of England annually from 1889 to 1910 (left axis), as well as the share of total sterling bills issued on the market (according to Nishimura's estimate) which were re-discounted by the Bank (right axis).

*Source:* Authors' computations on Bank of England Archives, *Comparative Statement of Discount Operations* (C30/3). Estimates of total sterling bills issued are from Nishimura, *Inland Bills*, p. 93.

**Figure 5.** Share of individual merchant banks in Gilletts' 1906 discounts and re-discounts to the Bank of England

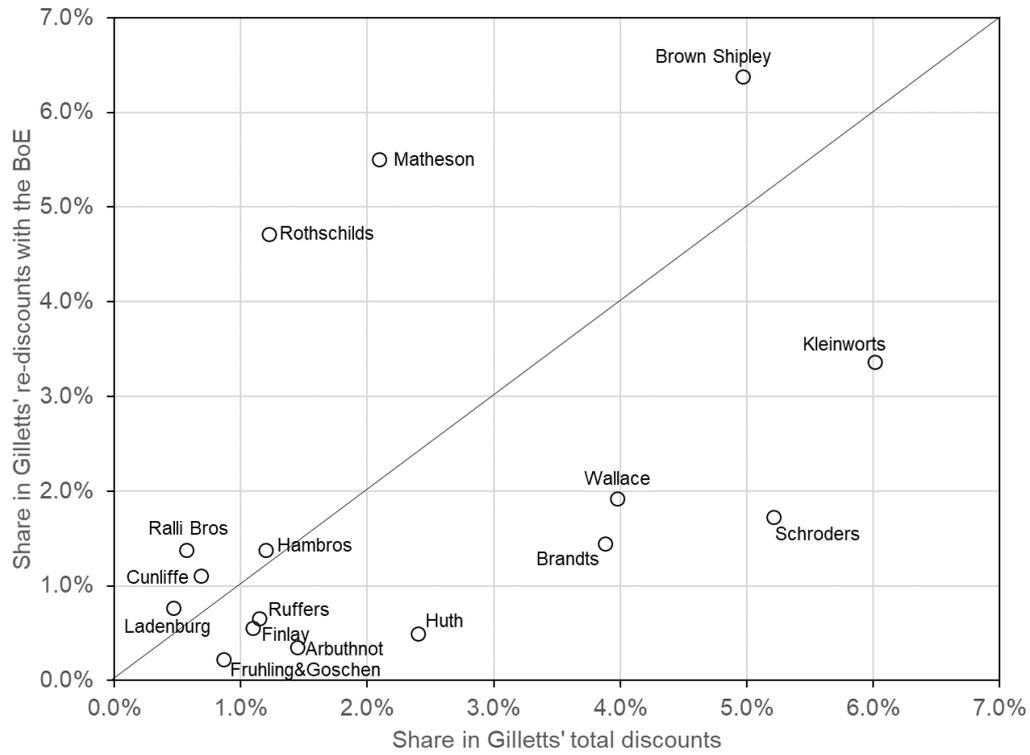

*Note:* The values on the horizontal axis show the share of various acceptors in Gilletts' total amount of bill discounts in the year 1906. The values on the vertical axis show the share of the same acceptors in the total amount of bills which Gilletts re-discounted to the Bank of England. Only merchant banks that are present in both samples are included in the figure.
*Source:* See text.

**Figure 6.** Geographical location of European borrowers (drawers) on the London bill market

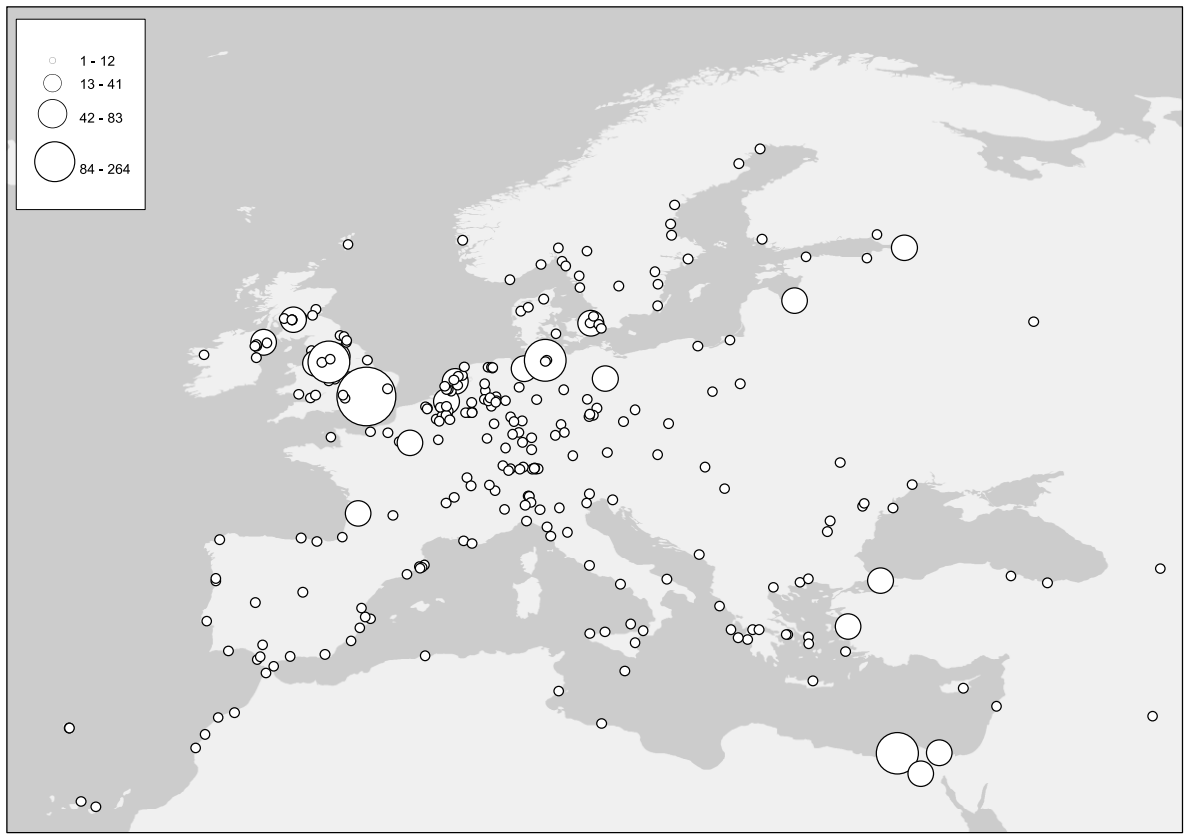

*Note:* This map shows the geographical location (at the city level) of all European drawers of sterling bills in our data set.
*Source:* Bank of England's *Discount Ledgers* (see text).

**Figure 7.** Dual structure of the accepting and discounting industries

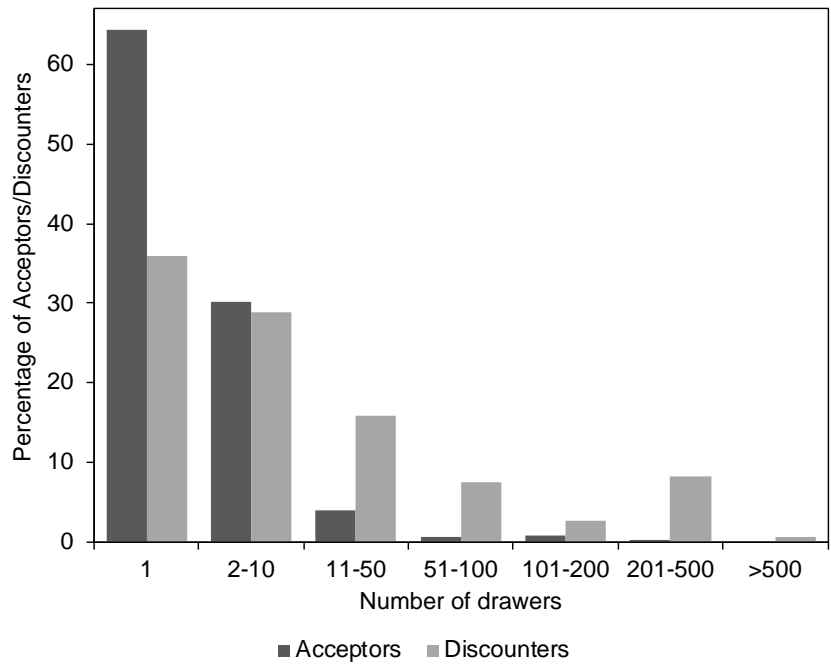

*Note:* This figure shows the frequency distribution of acceptors and discounters in terms of the number of drawers to which they were linked.
*Source:* Bank of England's *Discount Ledgers* (see text)

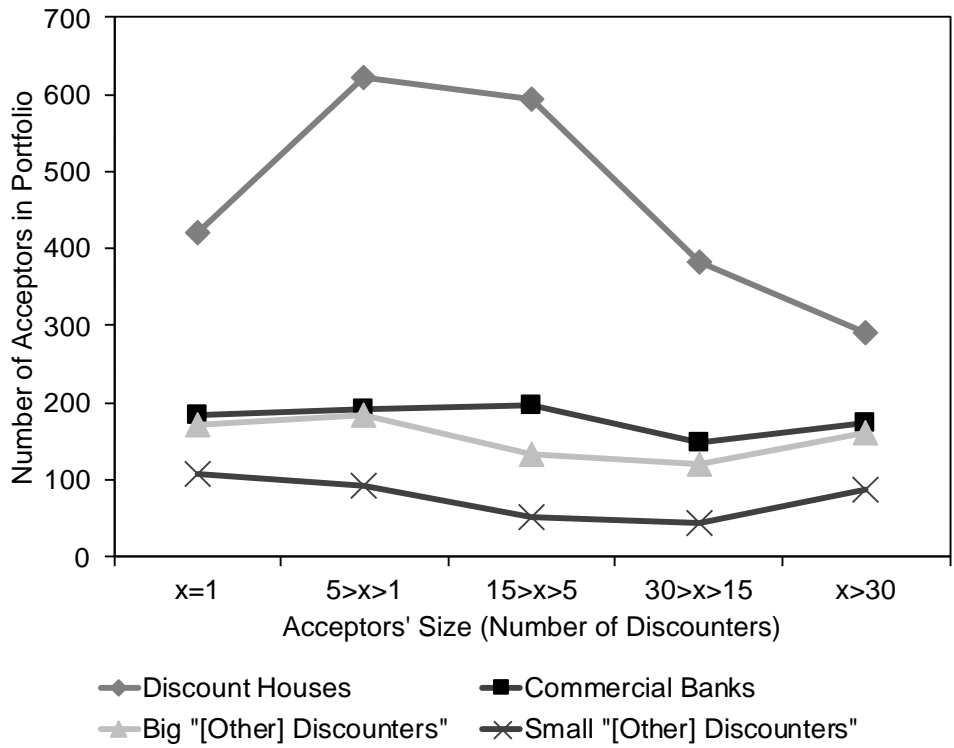

**Figure 8.** Discounters' portfolio of acceptors

*Note:* This figure describes the composition of the aggregate bill portfolios of different types of discounters (discount houses, commercial banks, big '[other] discounters', and small '[other] discounters') by acceptor size. Small '[other] discounters' are those linked to fewer than 18 acceptors, and big '[other] discounters' are those linked to at least 18 acceptors. The size of acceptors (x) is defined as their total number of discounters.

*Source:* Bank of England's *Discount Ledgers* (see text).

**Figure 9.** Discounters' portfolio of drawers

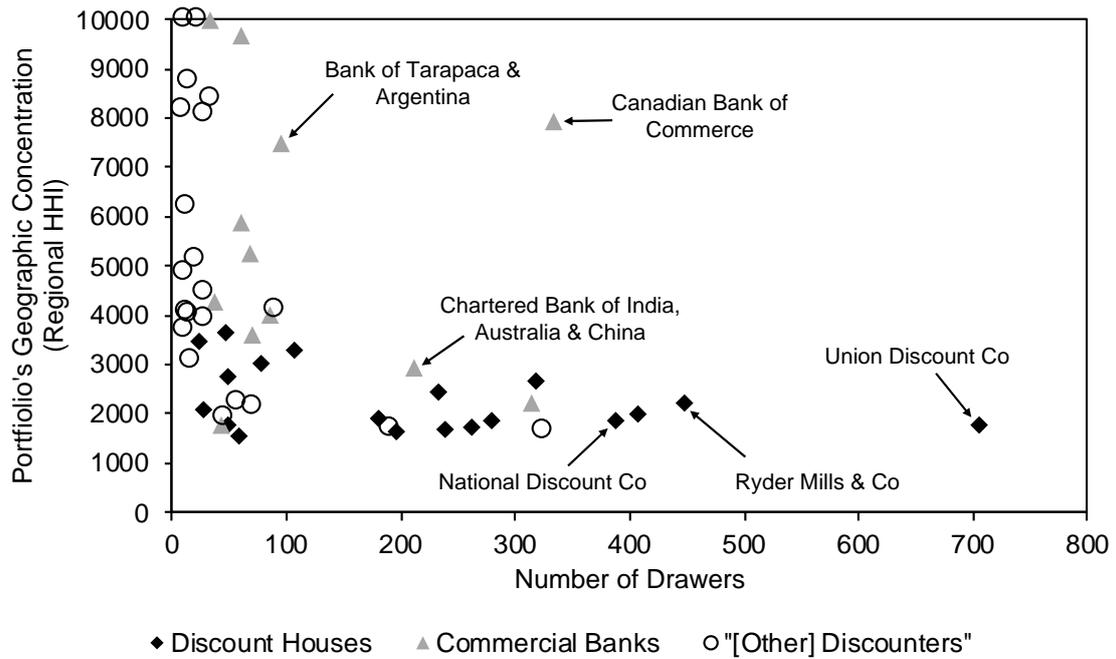

*Notes:* This figure plots the level of geographic concentration of each discounter's portfolio as a function of its size, defined as the number of drawers to whom the discounter is linked. The level of geographic concentration is assessed via the Herfindahl–Hirschman index, defined as the sum of the squares of the market shares of the nine regions in each portfolio; HHI values can range from 1,111 (perfect repartition among the nine regions) to 10,000 (perfect concentration in one region). The graph includes only those discounters (52 of the 145 in our data set) linked to at least 10 drawers. Regions are defined in the text.
*Source:* Bank of England's *Discount Ledgers* (see text).

**Figure 10.** Discount houses versus commercial banks' portfolios of drawers

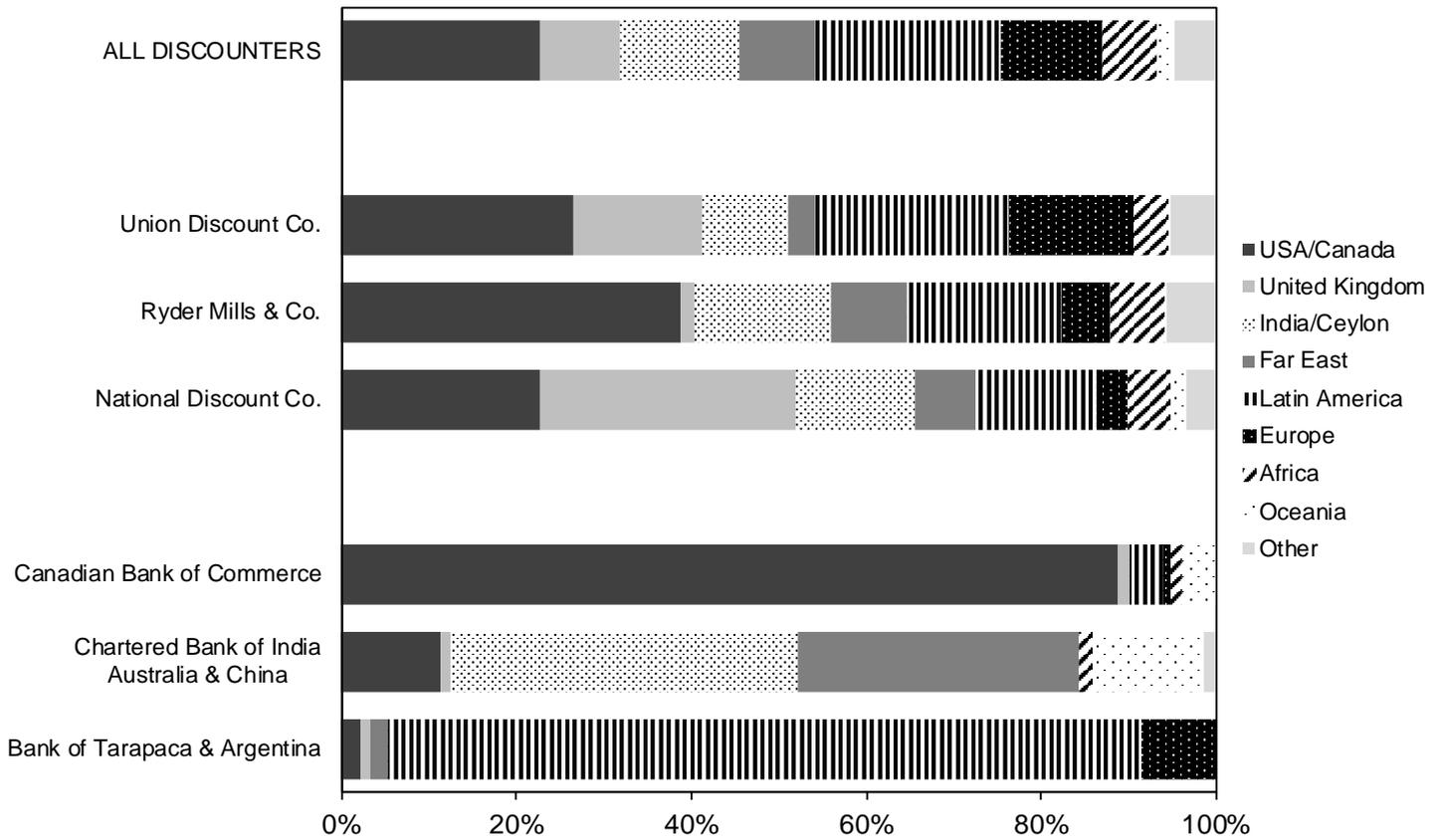

*Notes:* This figure illustrates the geographical location (via colour key) of the drawers of bills discounted by three discount houses and three Anglo-foreign commercial banks as well as the extent of their representation (via the *x*-axis percentages) in each institution's portfolio. Union Discount Co. was linked to 705 drawers, Ryder Mills & Co. to 447, National Discount Co. to 387, Canadian Bank of Commerce to 333, Chartered Bank of India Australia & China to 212, and Bank of Tarapaca & Argentina to 95.
*Source:* Bank of England's *Discount Ledgers* (see text).